\documentclass{IEEEtran}
\usepackage{cite}
\usepackage{amsmath,amssymb,amsfonts}
\usepackage{algorithmic}
\usepackage{graphicx}
\usepackage{textcomp}

\usepackage{color}
\usepackage{caption}
\usepackage{subfigure}
\usepackage{array}
\usepackage{tabularx}
\usepackage{multirow}
\usepackage{gensymb}

\usepackage{hyperref}
\usepackage{soul}
\usepackage{float}

\newcolumntype{L}[1]{>{\raggedright\arraybackslash}p{#1}}
\newcolumntype{C}[1]{>{\centering\arraybackslash}p{#1}}
\newcolumntype{M}[1]{>{\centering\arraybackslash}m{#1}}

\def\e{\begin{equation}}
\def\f{\end{equation}}
\def\_#1{{\bf #1}}
\def\@#1{_{\rm #1}}

\def\l#1{\label{eq:#1}}
\def\r#1{(\ref{eq:#1})}

\def\BibTeX{{\rm B\kern-.05em{\sc i\kern-.025em b}\kern-.08em
    T\kern-.1667em\lower.7ex\hbox{E}\kern-.125emX}}
	
\begin{document}
\title{Non-scattering Metasurface-bound Cavities for Field Localization, Enhancement, and Suppression}
\author{F.S.~Cuesta, \IEEEmembership{Member, IEEE}, V.S.~Asadchy, A.D.~Sayanskiy, V.A.~Lenets, 
M.S.~Mirmoosa, \IEEEmembership{Member, IEEE}, X.~Ma, S.B.~Glybovski,  and   S.A.~Tretyakov, \IEEEmembership{Fellow, IEEE}
\thanks{(Corresponding author: Francisco S. Cuesta)}
\thanks{F.S.~Cuesta, M.S.~Mirmoosa, and S.A.~Tretyakov 
are with the Department of Electronics and Nanoengineering, Aalto University, 
FI-00079 Aalto, Finland 
(e-mail: francisco.cuestasoto@aalto.fi; viktar.asadchy@aalto.fi;
 mohammad.mirmoosa@aalto.fi; sergei.tretyakov@aalto.fi).}
\thanks{V.S.~Asadchy was with the Department of General Physics, Francisk Skorina Gomel State University, 
246019 Gomel, Belarus. He is with the Department of Electronics and Nanoengineering, Aalto University
(e-mail: viktar.asadchy@aalto.fi).}
\thanks{A.D.~Sayanskiy, V.A.~Lenets, and S.B.~Glybovski 
are with the Department of Physics and Engineering, ITMO University, Russia
(e-mail: a.sayanskiy@metalab.ifmo.ru; vladimir.lenets@metalab.ifmo.ru; s.glybovski@metalab.ifmo.ru).}
\thanks{X.~Ma was with the Department of Electronics and Nanoengineering, Aalto University.
She is now with the Department of Electronics and Information, Northwestern Polytechnical University,
Xi'an, Shaan Xi, China
(e-mail: maxin1105@nwpu.edu.cn).}}

\maketitle

\begin{abstract}
We propose and analyse metasurface-bound invisible (non-scattering) partially open cavities where the inside field distribution can be engineered. 
It is  demonstrated both theoretically and experimentally that the cavities exhibit unidirectional invisibility   
at the operating frequency with enhanced or suppressed field at different positions inside 
the cavity volume. Several examples of applications of the designed  cavities are proposed and analyzed, in particular,   
cloaking sensors and obstacles,  enhancement of emission, and ``invisible waveguides''. 
The non-scattering mode excited in the proposed cavity is driven by the incident wave and resembles 
an ideal bound state in the continuum of electromagnetic frequency spectrum. In contrast to known bound states in the continuum, 
the mode can stay localized in the cavity infinitely long, provided that the incident wave illuminates the cavity. 
\end{abstract}


\section{Introduction}
\label{sec:introduction}
\IEEEPARstart{L}{ocalization} of electromagnetic waves plays central role for applications in various fields of technology such as lasers, filters, optical fibers, nonlinear devices. Especially in nanophotonics achieving high field concentration in optically compact systems is of paramount importance~\cite{Soref2006,Koenderink2015}. 
A simple example of wave localization is the echo inside a cave. Sound waves created by an object   inside the cave cannot escape outside and experience multiple reflections from its  hard walls before getting eventually absorbed. Meanwhile, the strength of sound   can be strongly amplified in this cavity due to the wave interference. However, waves which excite a resonant mode from outside are weakly coupled to the cavity and are mostly reflected back to the external space.  
Therefore, a pertinent question arises whether it is possible to localize waves in a space region without isolating it by walls and, most importantly, without disturbing the waves incoming into it from outside. Such space region would appear for an external observer as  free space (fully imperceivable), while it would operate as a resonant cavity for an observer inside it.

An apparent solution for realizing strong field localization in an open system is based on  Fabry-Perot interferometers~\cite{Fabry1899,Perot1899,Culshaw1959} formed by two  partially transparent mirrors separated by a specific distance.  At  discrete wavelengths, lossless Fabry-Perot resonators pass through all the power carried by the incident waves due to the destructive interference of reflected waves between the mirrors. 
However, these resonators are still perceivable for an external observer since they alter the phase of the transmitted waves, that is, they produce scattered waves. This forward scattering represents little-or-no inconvenience for applications that require unitary power  transmission. However, performance of synchronous systems, such as array beamformers \cite{Godara1985}, homodyne interferometry \cite{Dandridge1982}, and high-resolution astronomical observations \cite{akiyama2019first}, is degraded drastically in the presence of   forward scattering. Hence, solution for latter scenarios requires open non-scattering cavities which the field outside of the cavity is not modified at all. Furthermore, studies of non-scattering resonant objects are of fundamental interest as realizations of bound eigenstates in the continuum.  

Theory of non-scattering and non-radiating bodies has a long history dating back to the work by Ehrenfest, where he recognized that radiationless current distributions are  possible~\cite{Ehrenfest1910,Kerker1975,Devaney1978,Gbur2003}, being the starting point for new and innovative concepts and applications \cite{Miroshnichenko2015}.
Nevertheless, such systems do not necessary allow one to achieve strong fields localizations. For example, consider a non-scattering configuration which includes two  planar metasurfaces~\cite{Epstein2016, Elsakka2016}: Each sheet does not produce back scattering, while in the forward direction their scattering can be mutually compensated. As a result, the field between the interfaces is identical to that of the incident wave and cannot reach extreme values.

\begin{figure*}[tb]
 \center
  \includegraphics[width=0.96\textwidth]{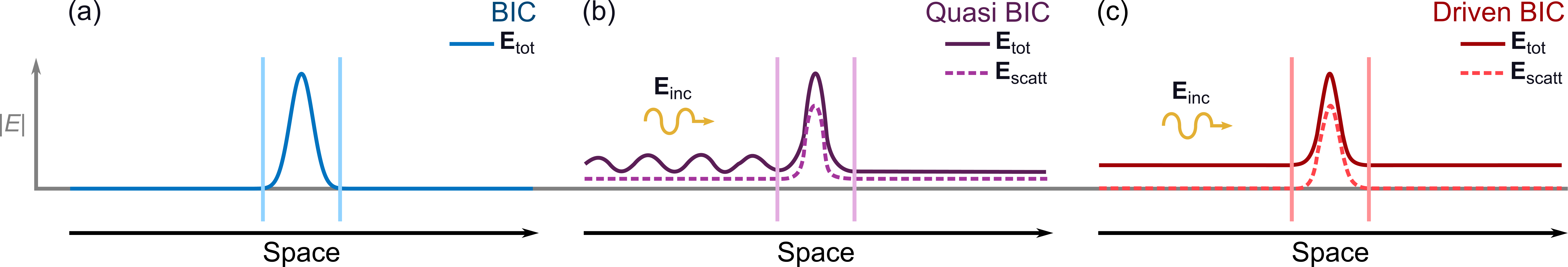}
  \caption{Conceptual illustration of different states in the continuous spectrum of frequencies. (a) Bound state in the continuum (BIC). Such a state or eigenfunction is fully localized in a finite region. Once the state is excited, it will be  bound for an infinite time.   (b) Quasi BIC corresponding to an eigenfunction which locally resembles BIC but extends outside the region. Once excited, it slowly leaks out through radiating waves, even at the presence of an incident wave. Total field in shown in solid lines, while the field produced by scattered waves is shown in dashed lines.  (c) Driven BIC which exists during the time when the region is illuminated by an external source (driven by the incident wave). The state is perfectly bound to the  region and can exist even in structures with a finite quality factor.  }
  \label{fig:BIC_comp}
\end{figure*}

Thus, realization of invisible cavities implies combination of field enhancement properties of Fabry-Perot resonators and  invisibility features of \emph{non-scattering} electromagnetic systems. 
This scenario to some extent  resembles  bound states in the continuous spectrum of frequencies (BICs), i.e. \emph{non-radiating} eigenmodes captured into a localized volume~\cite{Hsu2016} [see Fig.~\ref{fig:BIC_comp}(a)].
First studies of this concept were made by von Neumann and Wigner \cite{ vonNeumann1929, Stillinger1975} in quantum mechanics, and expanded through the years to other branches of physics:  photonics~\cite{Marinica2008,Moiseyev2009,Plotnik2011,monticone_embedded_2014,silveirinha_trapping_2014,lannebere_optical_2015,liberal_nonradiating_2016,Sadrieva2016,Rybin2017,Bulgakov2018,Carletti2018,Koshelev2018,Koshelev2018a,Krasnok2018,Liu2018}, acoustics~\cite{Bortolani1983,Linton2007}, electronics~\cite{Albo2012,Zhen2014}, and others~\cite{Friedrich1985}.
Ideal BICs, with an infinite quality factor, cannot be excited from outside and are practically impossible due to dissipation loss present in all physical systems. Instead,  it is more relevant to speak about  quasi BICs,  with a large but finite lifetime, which are feasible under realistic conditions [see Fig.~\ref{fig:BIC_comp}(b)]. 
A quasi BIC, in the absence of incident waves, slowly leaks the stored energy away from the localization region. Due to that, an external incident wave is used to extend the lifetime of the eigenmode, compensating the leaked energy. In the best scenario, back scattering is eliminated but forward scattering shifts the phase of the transmitted wave. 
In contrast to quasi BIC,   the eigenmode excited inside invisible resonators which we introduce and study here [see Fig.~\ref{fig:BIC_comp}(c)] does not decay and at the same time does not scatter under the presence of an incident wave. In fact, the scattering profile of an invisible resonator resembles the field distribution of an ideal BIC [compare the dashed line of Fig.~\ref{fig:BIC_comp}(c) with BIC profile of Fig.~\ref{fig:BIC_comp}(a)], resulting in a cavity eigenmode ``\emph{driven}'' by  the incident wave.
Their distinctive features make invisible cavities good candidates for various applications, while practical implementation of quasi BICs  have been scarce to date and limited to very recent works~\cite{Kodigala2017, Romano2018,Ha2018}.

In this paper, we propose and design   invisible cavities formed by two parallel thin patterned metal sheets (1D scenario). We call the sheets planar metasurfaces~\cite{Glybovski2016,Yu2014} (or frequency-selective surfaces in special cases~\cite{Munk2005}).
We demonstrate, both numerically and experimentally, unidirectional invisibility  of the cavity at the operating frequency with  localized field enhancement and suppression at different positions between the metasurfaces. The amplitude of the field  within the cavity depends on the degree of transparency  of the metasurfaces, and, in the limiting case, the non-scattering modes (so-called driven BICs) converge to ideal BICs. Several examples of applications of the designed invisible cavities are proposed and analyzed, in particular,   cloaking sensors and obstacles. Furthermore, we demonstrate that in the extreme reactances limit, the proposed cavities exhibit essentially different convergence to the ideal BICs as compared to conventional cavities, providing an efficient route towards realizations of finite-time excitation of BICs with theoretically infinite lifetime. 


\section{Design of invisible cavities}

\begin{figure}[!h]
    \centerline{\includegraphics[width=70mm]{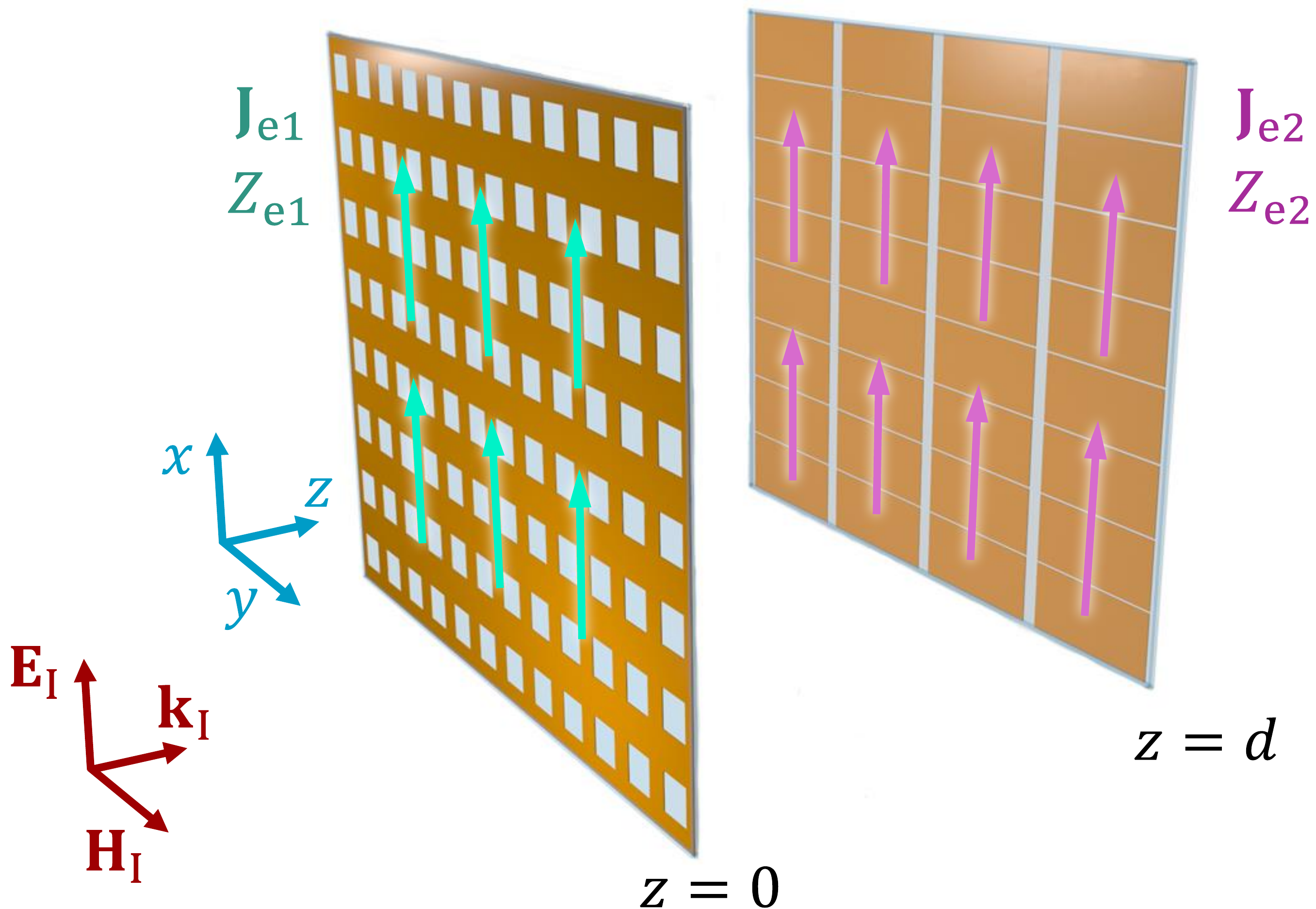}}
    \caption{Geometry of the problem. Incident plane wave illuminates a cavity formed by two metasurfaces.  \label{fig:two_EP_fields} }
\end{figure}

Let us consider a one-dimensional scenario, where two metasurfaces infinite in the $xy$ plane are separated by a distance $d$ in a homogeneous medium (assumed to be vacuum in this work), as shown in Fig.~\ref{fig:two_EP_fields}.
The metasurfaces represent artificial composite sheets: subwavelength periodic structures  with negligible thickness, designed  for   electromagnetic wave manipulations~\cite{Glybovski2016,Yu2014}. 
When such a cavity  is excited by a normally incident plane wave, electric  currents (characterized by the surface current densities $\_J_{\rm e1}$ and $\_J_{\rm e2}$) are excited in the metasurfaces, which produce   scattered waves inside and outside the cavity. 
Each metasurface is characterized by its grid impedance ($Z\@{e1}$ and $Z\@{e2}$, respectively), defined as the  ratio of the tangential component of the surface-averaged  electric field  at the metasurface position  and   the induced electric current density.

We design the cavity so that it produces  no scattered waves outside it, meaning that  the reflected wave should be eliminated ($E\@{R}=0$) and the transmitted wave should be equal to the incident wave ($E\@{T}=E\@{I}$) without any phase lag. 
Field solutions on the base of the impedance boundary conditions at both sheets (see Appendix A) lead us to the following conditions on the grid impedances and the distance $d$:
\begin{equation}
 Z\@{e1}=-Z\@{e2}, \quad d=\frac{n \lambda_{\rm op}}{2}.
 \label{eq:eqfreespace}   
\end{equation}
We see that the cavity is invisible only when the distance between the metasurfaces is fixed to an integer of the half-wavelength of the supporting medium. On the other hand, the metasurfaces can have arbitrary complex grid impedances as long as the two values satisfy Eqs.~\eqref{eq:eqfreespace}.

This relation between the grid impedances implies that if one metasurface is lossy, then the second one should be pumped with external energy to compensate the power dissipated in the lossy sheet. In the limiting case of active/lossy metasurface combination, when the impedances $Z\@{e1}=R\@{e}$ and $Z\@{e2}=-R\@{e}$ are purely real, the cavity behaves as a teleportation slab with parity-time symmetry~\cite{Radi2016}. However, for practical implementations of the structure, it is desirable to use purely imaginary grid impedances (using easily realizable capacitive and inductive metasurfaces), which is the case when   dissipation is negligible (influence of inevitable small losses on the cavity performance is elucidated below).


While the scattering from such invisible cavities is completely suppressed, the fields inside it are different from the incident field and can be found as combinations of forward and backward propagating waves. This solution corresponds to the homogenized impedance model of metasurfaces, which is valid when the distance between the sheets is considerably larger than the metasurface array period (so that the evanescent waves excited due to the discrete structure of the metasurface decay to negligible level). In this case, the field inside the cavity can be modelled as a standing wave (see derivations in Appendix~B).
\begin{figure}[tb]
    \centerline{\includegraphics[width=80mm]{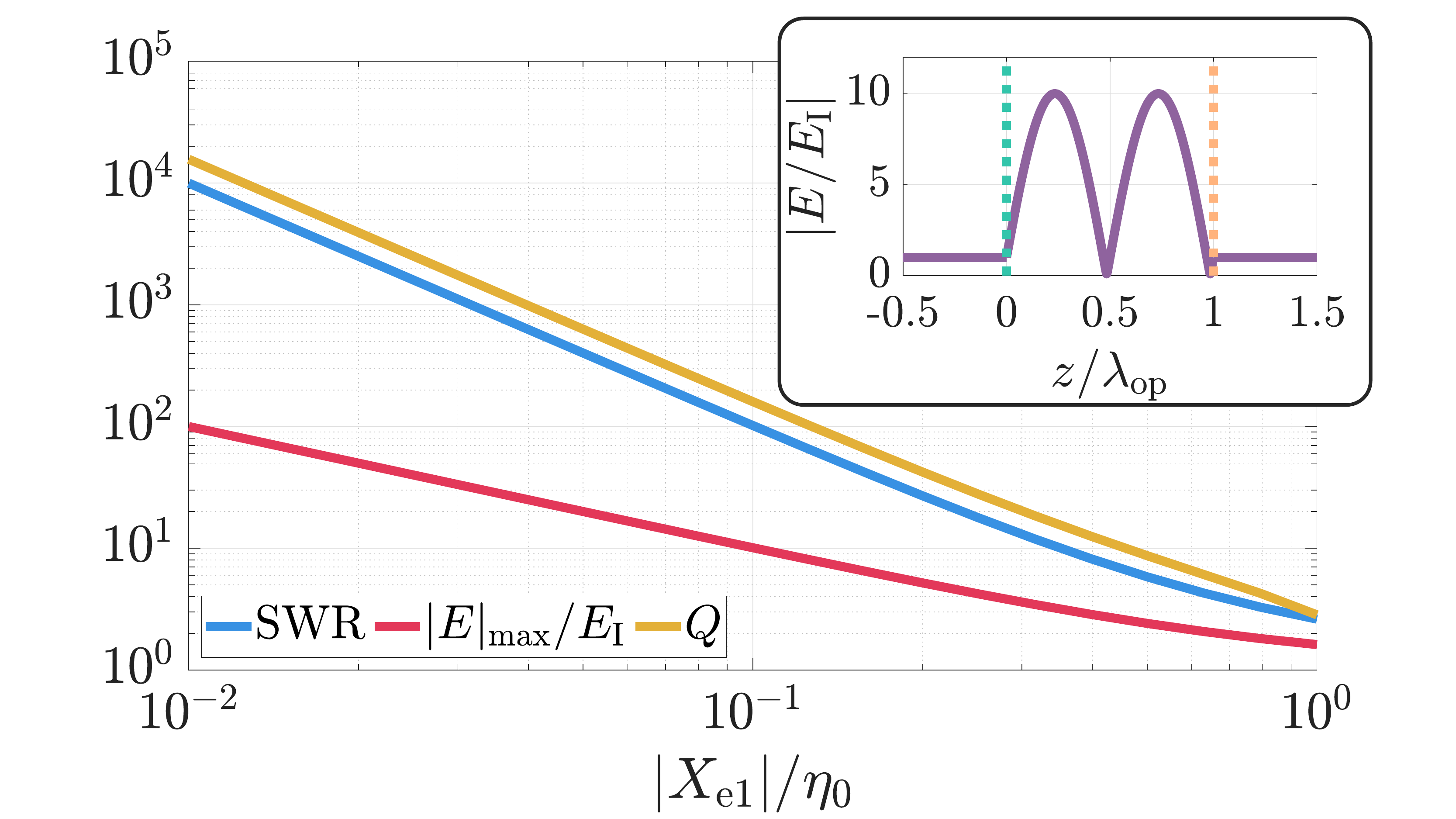}}
    \caption{Standing wave ratio,  normalized electric field maximum, and quality factor $Q$ for an invisible cavity  with the thickness $d=\lambda_{\rm op}$. The inset shows the electric field distribution across the cavity for the case $X\@{e1}\approx 38\,\Omega$ (${\rm SWR}=100$).}
    \label{fig:teo_performance}
\end{figure} 
Therefore, the ratio between the maximum and minimum values of the electric field magnitude inside the cavity can be characterized by the standing wave ratio (SWR), which is defined for the invisible cavity as
\e {\rm SWR}=\dfrac{\left\vert E\right\vert\@{max}}{\left\vert E\right\vert\@{min}}=\dfrac{\vert 2Z\@{e1}-\eta_0\vert+\eta_0}{\vert 2Z\@{e1}-\eta_0\vert-\eta_0}.\label{eq:SWR}\f
Figure~\ref{fig:teo_performance} illustrates the dependence of the standing wave ratio for lossless invisible cavities versus the grid reactance. 
Notice that SWR increases as both grid impedances converge to zero. This is important as highly localized fields are achieved with metasurfaces whose properties are close to those of a continuous sheet of perfect electric conductor (PEC).

Another useful characteristic of  cavity properties is the quality factor $Q$, which measured the relative bandwidth of the transmission window: $Q=\omega\@{c}/\Delta \omega$. By defining the relative bandwidth by the half-power level, we obtain the following expression for a lossless invisible cavity with near-PEC metasurfaces (see derivation in Appendix~B): 
\e Q\approx n\pi \left[ 4 \left(\dfrac{X\@{e1}}{\eta_0}\right)^2\sqrt{1-4 \left(\dfrac{X\@{e1}}{\eta_0}\right)^4}\right]^{-1}. \label{eq:qfactor_redux}\f
The   quality factor grows to infinity when the grid impedances tend to zero (the PEC limit) approximately as $1/X\@{e1}^2$, and it  is  also directly proportional to the distance between the metasurfaces (represented by the integer $n$). 
Thus, maximization of the quality factor, desired for many applications, is achieved when the cavity thickness is large and the grid impedance is small, as shown in Fig.~\ref{fig:teo_performance}. It should be noted that this definition of the   quality factor implies lossless and dispersionless unit cells of the metasurfaces (additional frequency dispersion of lossless metasurfaces generally results in increasing the  quality factor due to additional storage of reactive energy). 

\section{Driven bound states in the continuum}
As it was mentioned above, for finite values of the sheet reactances the non-scattering eigenmode (eigenstate) of invisible metasurface-bound cavities is a driven bound state in the continuous spectrum of frequencies, in contrast to  full-transmittance modes of conventional Fabry-Perot resonators.
At the  frequency of the mode, the scattered field is completely localized inside the cavity [the cavity appears invisible as is illustrated in Fig.~\ref{fig:BIC_comp}(c)], while at all higher and lower frequencies the cavity produces scattered fields, yielding a continuous spectrum of unbounded states.
The name ``driven'' comes from the fact that this bound state does not leak energy only when the cavity is illuminated by a plane wave.
If the illumination is switched off, the excited mode energy will eventually leak away. 

\begin{figure}[bt]
  \centerline{\includegraphics[width=80mm]{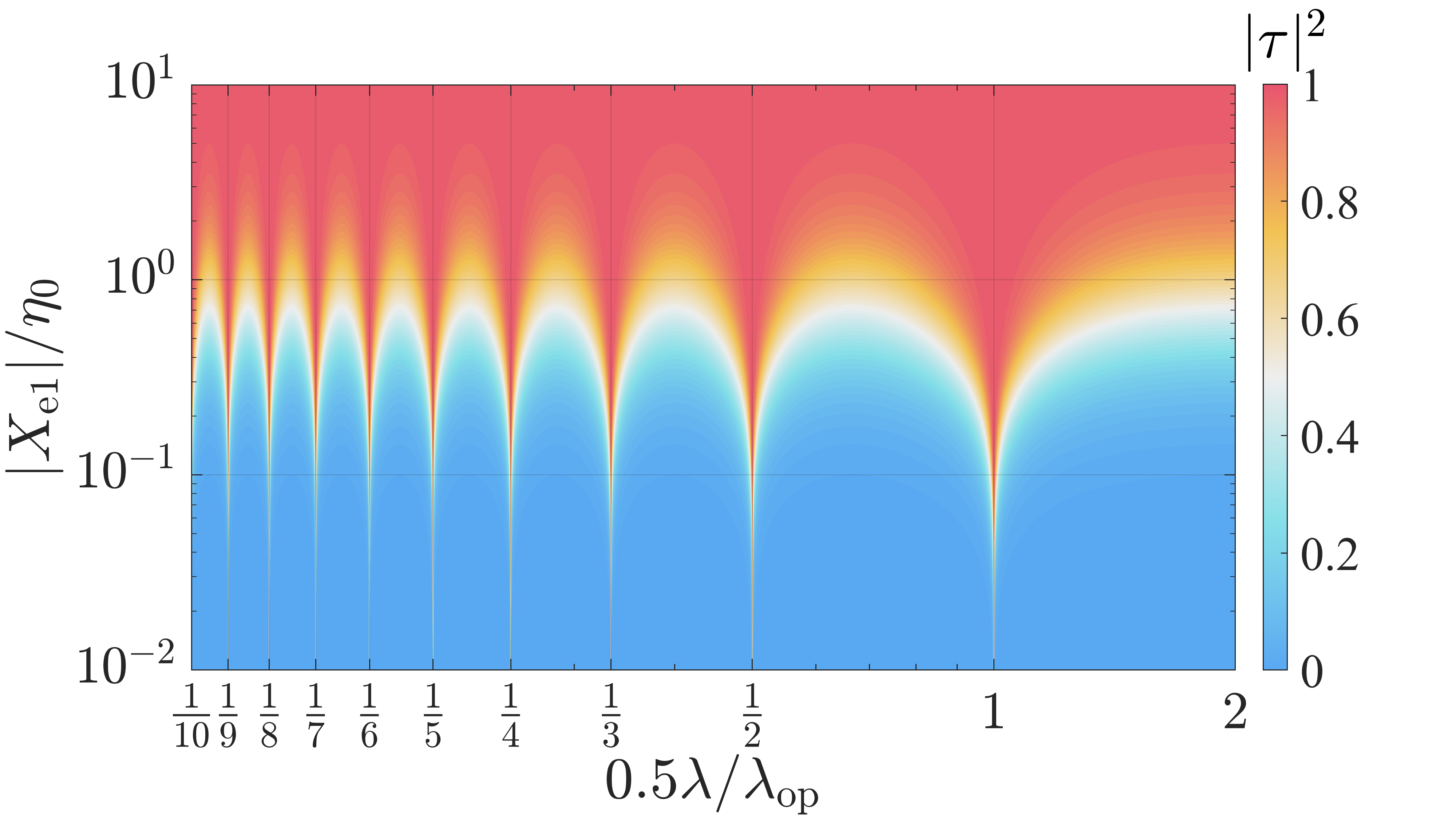}}
     \caption{Transmittance through the cavity as a function of the metasurface grid reactance   and the wavelength of incident waves.  At wavelengths $\lambda=2\lambda\@{op}/n$, the cavity approaches ideal BIC   in the limit of $X_{\rm e1} \rightarrow 0$. In this example, the distance between  the two  metasurfaces is $d=\lambda\@{op}$.}
    \label{fig:Frequency_transmission}
 \end{figure}

A distinctive feature of driven BICs is that they have infinite lifetime (limited only by  duration of the external illumination) in cavities with a finite Q-factor. According to Fig.~\ref{fig:teo_performance}, as the grid impedances of the metasurfaces approach zero, the Q-factor tends to infinity, and the driven eigenmode becomes an ideal BIC. Figure~\ref{fig:Frequency_transmission} shows the transmittance through the cavity as a function of the metasurface grid reactance $X_{\rm e1}$ and the wavelength of incident waves $\lambda$. One can clearly recognize sharp resonances of the cavity at $\lambda=2d/n$ ($n$ is a positive integer) corresponding in the limit of $X_{\rm e1} \rightarrow 0$ to ideal BICs.

\begin{figure}[bt]
    \centerline{\includegraphics[width=80mm]{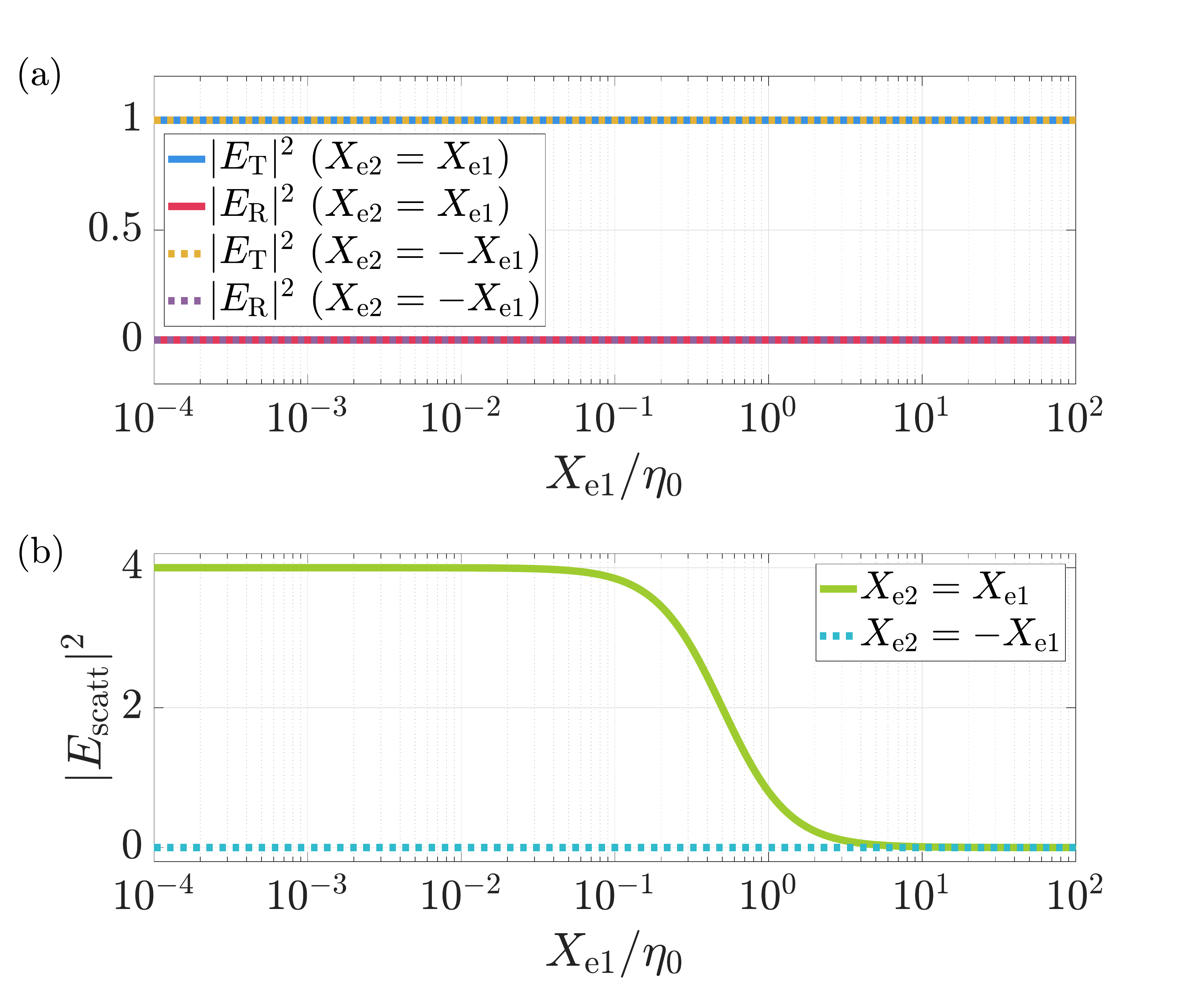}}
    \caption{(a) Reflectance and transmittance through resonant cavities with $X_{\rm e1}=-X_{\rm e2}$ and $X_{\rm e1}=X_{\rm e2}$. The first scenario corresponds to an invisible cavity, and the second one to a Fabry-Perot resonator tuned for unitary transmission. In both scenarios, an incident wave with unitary electric field $E\@{I}=1$  is assumed.  (b) Comparison of the forward scattering produced behind the cavity. In the limit of zero reactances, the two scenarios provide drastically different response,  although the limiting values of $X_{\rm e1}$ and $X_{\rm e2}$ are the same.}
    \label{fig:AS_TF_Emax}
\end{figure}

Interestingly, in the limit of $X_{\rm e1}\rightarrow 0$ and $X_{\rm e2} \rightarrow 0$, both metasurfaces become identical. Nevertheless, it is important that the grid reactances $X_{\rm e1}$ and $X_{\rm e2}$ tend to zero from the opposite sides so that condition $X_{\rm e1}=-X_{\rm e2}$ holds.
Figure~\ref{fig:AS_TF_Emax}(a) shows reflectance and transmittance through the cavity for two convergence  scenarios: When $X_{\rm e1}=-X_{\rm e2} \rightarrow 0$ and when $X_{\rm e1}=X_{\rm e2}  \rightarrow 0$.

In the former case, the cavity is invisible for all reactance pairs; while in the latter case, the FP-based cavity remains transparent everywhere.  To quantify the degree of invisibility we plot the difference between the transmitted and the incident wave amplitudes,  $\_E\@{scat}=\_E\@{T}-\_E\@{I}$ on Fig.~\ref{fig:AS_TF_Emax}(b). The dramatic difference of the two curves is evident. We see that the Fabry-Perot cavity formed by two identical metasurfaces becomes invisible only in the trivial limit of infinite sheet reactances (no metasurfaces at all). For moderate and small values of the sheet reactance, scattering is strong although the power transmittance tends to unity for small reactances. Thus, while invisible cavities do not scatter at all, Fabry-Perot resonators produce a strong forward scattered wave that, combined with the incident wave, results in a phase shift of the transmitted wave.


\section{Realization of invisible cavities}
In order to verify the theoretically predicted behavior of invisible cavities, a set of waveguide experiments was carried out. It is well known that the fundamental TE$_{10}$ mode propagating in a waveguide with the rectangular cross-section of the dimensions $a\times b$ with electric field magnitude $E_0$ and the wavenumber $k_z=\sqrt{k_0^2-\left(\frac{\pi}{a}\right)^2}$ 
 can be described as a superposition of two TEM plane waves as follows:
\begin{equation} 
E_y(x,z)=A e^{jk_0\cos\theta z} \left( e^{jk_0\sin\theta x} - e^{-jk_0\sin\theta x} \right).
\label{TE10}
\end{equation}
In the last expression $\theta=\cot^{-1}\sqrt{\left(\frac{k_0 a}{\pi}\right)^2-1}$ is the frequency-dependent angle of incidence of the plane waves with respect to the $z$-axis of the waveguide and $A=- jE_0/2$ is their magnitude.
Therefore, the problem of scattering of the fundamental mode by a thin obstacle placed in the cross-section of the waveguide is electromagnetically equivalent to that of plane-wave scattering by a periodic array of such obstacles with the periodicities $a$  and $b$ along the $x$ and $y$ axes, respectively.
Although this method does not provide an arbitrary choice of the incidence angle at the operational frequency and does not support normal incidence, it is still applicable to analyze the proposed invisible cavities. 
As the wavelength of the TE$_{10}$ mode equal to $\Lambda=2\pi/k_z$ differs from the wavelength in free space, in our approach the separation of metasurfaces in the cavity (multiple of half-wavelengths) should be modified accordingly. 
It is, however, possible to demonstrate that the requirement of mutually complex-conjugate grid impedances still holds for the metasurfaces providing theoretically perfect non-scattering regime in the case of oblique incidence. 

%
\begin{figure*}[tbh]
  \centering
  \includegraphics[width=1\linewidth]{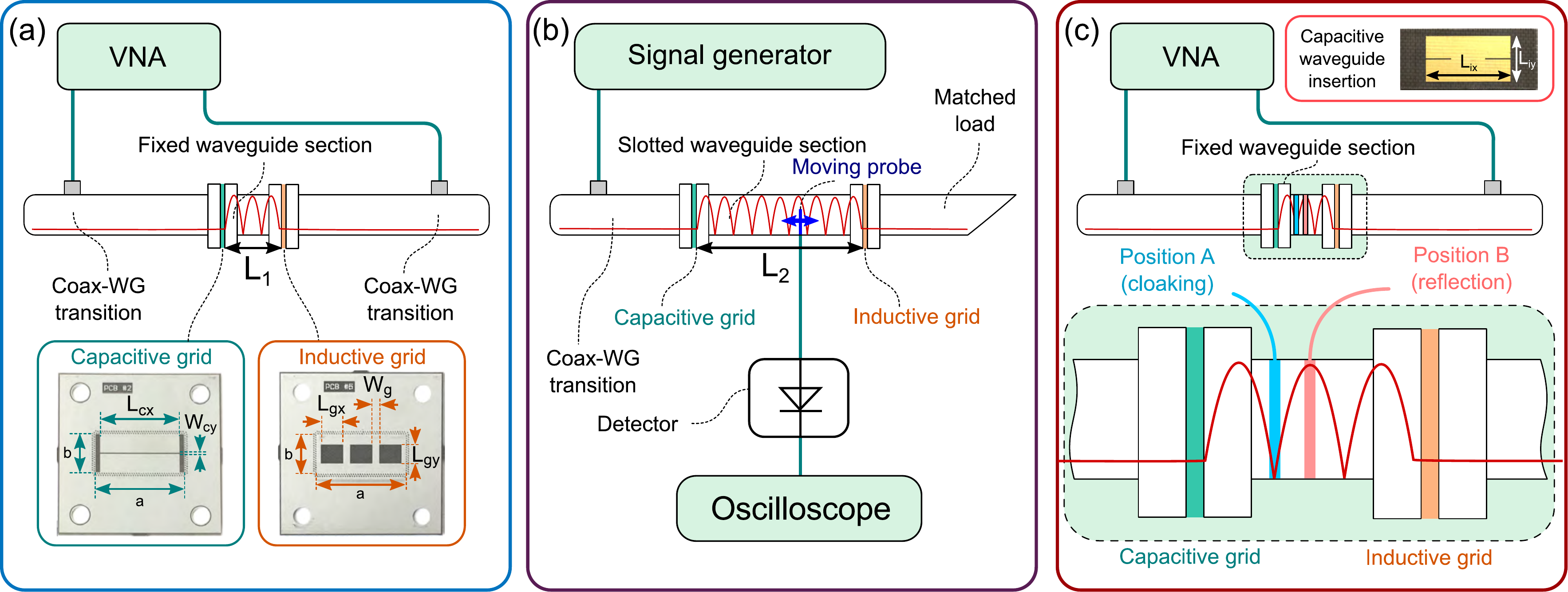} 
  \caption{Schematics of the setup used to demonstrate the properties of the proposed invisible cavities. The cavity is composed of  unit cells (printed on a circuit board) of capacitive and inductive metasurfaces placed into a WR-90 waveguide section. The subplots illustrate the three performed experiments: (a) Setup for measurements of transmission and reflection coefficients of the invisible cavity (insets show the manufactured metasurface unit cells of the cavity); (b) setup for studying the field distribution in the invisible cavity; (c) setup for analyzing scattering suppression (Position A) and  enhancement (Position B) from the object inserted inside the cavity (shown in the up-right corner).}
  \label{Setup}
\end{figure*}

To reach the non-scattering regime, both the capacitive and inductive metasurfaces,  represented by their single rectangular unit-cells, were numerically optimized to provide the grid impedance of  $\pm j 31\,\Omega$. The capacitive metasurface unit cell comprised a rectangular copper patch of the length $L_{\text{c} x}=20.7$~mm and height $b-W_{\text{c} y}=9.7$~mm (here $W_{\text{c} y}=0.3$~mm is the gap width)  printed on a 0.5-mm-thick Arlon AD250 substrate with the relative permittivity of 2.5 and the dielectric loss tangent 0.0018 [see illustration in Fig.~\ref{Setup}(a)]. To ensure the grid impedance of $1/(j\omega C)= -j 31~\Omega$ at $f_1=11.11$~GHz, the width of the gap between adjacent patches in the corresponding periodic metasurface $W_{\text{c}x}$ was parametrically tuned to 0.3~mm. The unit cell of the inductive metasurface was represented by three rectangular apertures of the dimensions $L_{\text{g}x} \times L_{\text{g}y} = 4.6 \times 5$~mm$^2$ each separated by copper strips of the widths $W_{\text{g}}=3$~mm printed on a similar substrate. The latter value was chosen to provide the required value of the grid impedance of $j\omega L = j31~\Omega$. The parameters of $W_{\text{cx}}$ for the capacitive unit cell and $W_{\text{g}}$ for the inductive one were determined separately by placing the corresponding structure into a waveguide section terminated by two waveguide ports and extracting the grid impedances from measured reflection and transmission coefficients. Due to the losses in the copper metallization and the  substrates, both grid impedances have a small real part. The extracted impedance for the capacitive grid was $Z_{\text{c1}}=(0.17-j31)~\Omega$ and $Z_{\text{c2}}=(0.15+j31)~\Omega$ for the inductive one. These real parts of the grid impedances were taken into account when comparing the experimental results with numerical data (consider information provided in Appendix~C).
\begin{figure*}[tb]
  \centering
  \includegraphics[width=1\linewidth]{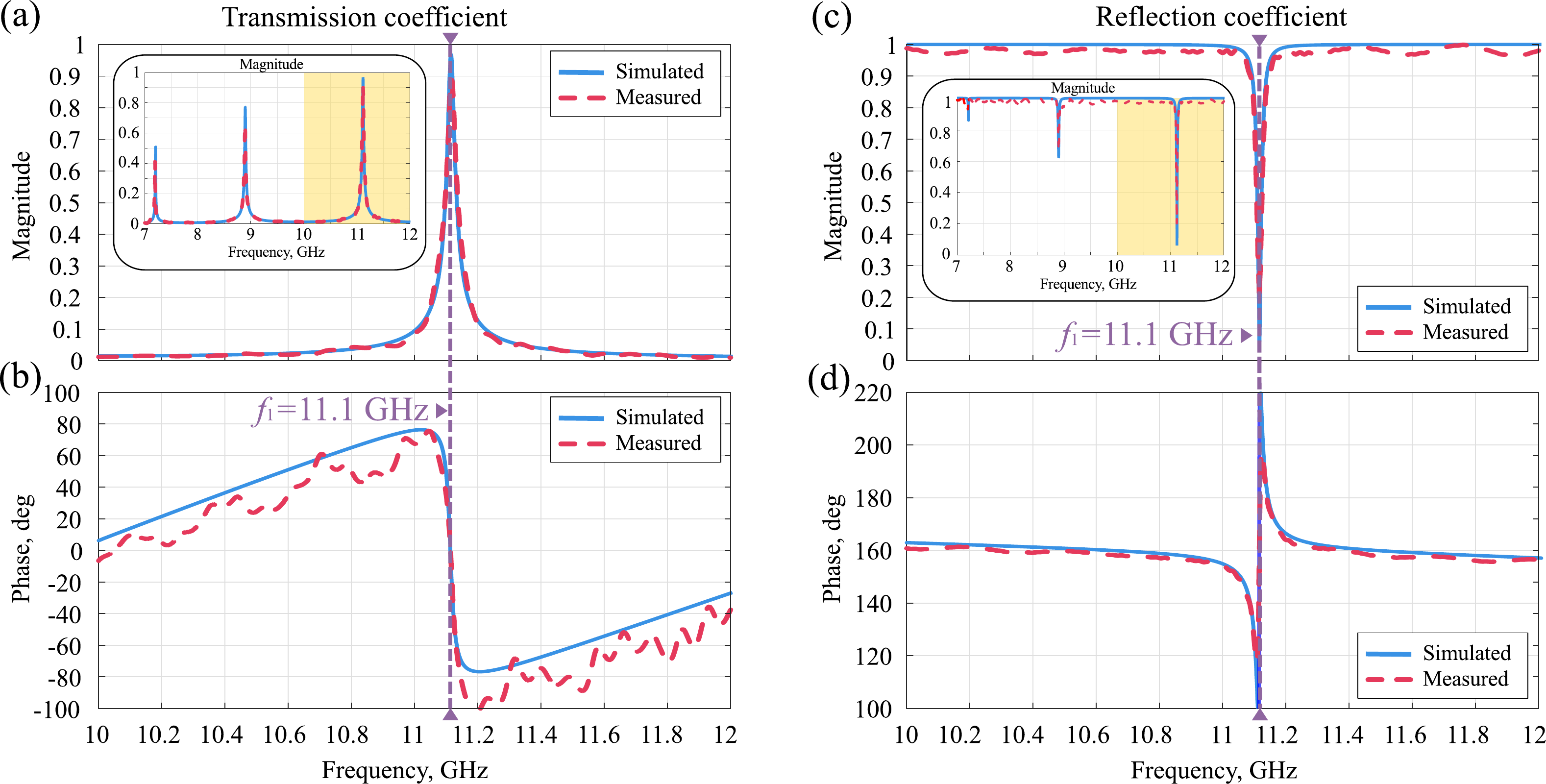} 
  \caption{Simulated and measured magnitudes and phases of the transmission (left column) and reflection (right column) coefficients in the frequency range from $10$ to $12$~GHz (not power coefficients). The insets show the transmission and reflection coefficient magnitudes versus frequency in the extended range from $7$ to $12$~GHz.}
  \label{T_R_WG50}
\end{figure*}

The aim of the first experiment was to demonstrate the operation of the invisible cavity or, in other words, to
show that properly choosing  the properties of the metasurfaces, high transmission at the resonance is achieved along with zero transmission phase.
In the waveguide setup, each of the two unit cells was surrounded with a continuous metalized area on both layers of the PCB with multiple vias at the periphery of the waveguide cross-section to ensure no current discontinuity on the waveguide walls. The photographs of the manufactured unit cells to be inserted before and after the waveguide section are shown in insets of Fig.~\ref{Setup}(a). 
The distance between them is three half-wavelengths, i.e. $L_1=3\Lambda/2$.
The complex transmission and reflection coefficients were retrieved from S-parameters measured by a vector network analyzer (VNA) using additional calibration data corresponding to an empty waveguide and a metal plate connected to the waveguide instead of each metasurface unit cell.
The experimental setup used  for characterization of the reflection and transmission coefficients of the invisible cavity is schematically shown in Fig.~\ref{Setup}(a). It is based on a uniform WG-90 waveguide section fixed between two coax-to-waveguide transitions from Agilent X11644A WR-90 X-band kit connected to ports of a vector network analyzer  Agilent E8362C. The length $L_1$ of the section was equal to $50$~mm, which for the waveguide cross-section of $a\times b = 23 \times 10$~mm$^2$ resulted in the third thickness  resonance condition at $f_1=11.11$ GHz. In other words, $L_1$ equals to three half-wavelengths in the waveguide at $f=f_1$, while the wavelength $\Lambda$ equals to $33.35$~mm.

Figure~\ref{T_R_WG50} shows  the measured transmission (left) and reflection (right) coefficients through the cavity in the frequency range from $10$ to $12$~GHz.
The insets show the same parameters but in the extended frequency range from $7$ to $12$~GHz. As one can see from the plots, there are three separate thickness resonances in the extended range. At each resonant frequency, the transmission coefficient magnitude is high, however it exceeds 0.9 only at the third resonance ($11.11$~GHz), i.e. at the frequency where the impedances of the inserted metasurfaces were designed to be complex conjugate of each other. The experimentally achieved transmission coefficient magnitude at the maximum was $0.9$ instead of $0.97$ predicted by the simulation due to some misalignment of the capacitive and inductive unit cells with respect to the cross-section plane. These results are in good agreement with theoretical predictions for lossy metasurfaces  (see Appendix~C). Indeed, the theoretical data provide estimated transmission coefficient of $\sqrt{0.9}\approx 0.95$ and nearly zero reflection coefficient for the designed metasurfaces with $R_{\rm loss}/|X_{\rm e}| \approx 5\cdot 10^{-3}$.
Furthermore, as one can check from the measured phase plots in Fig.~\ref{T_R_WG50}(b), the transmission coefficient phase crosses  zero at the same resonance frequency. The measured quality factor is $Q=414$.
Therefore, it is experimentally confirmed that the cavity is indeed practically invisible (non-scattering) for the chosen excitation. The experimental curves are in good correspondence with numerically calculated ones. 
\begin{figure}[bt]
  \centering
  \includegraphics[width=1\linewidth]{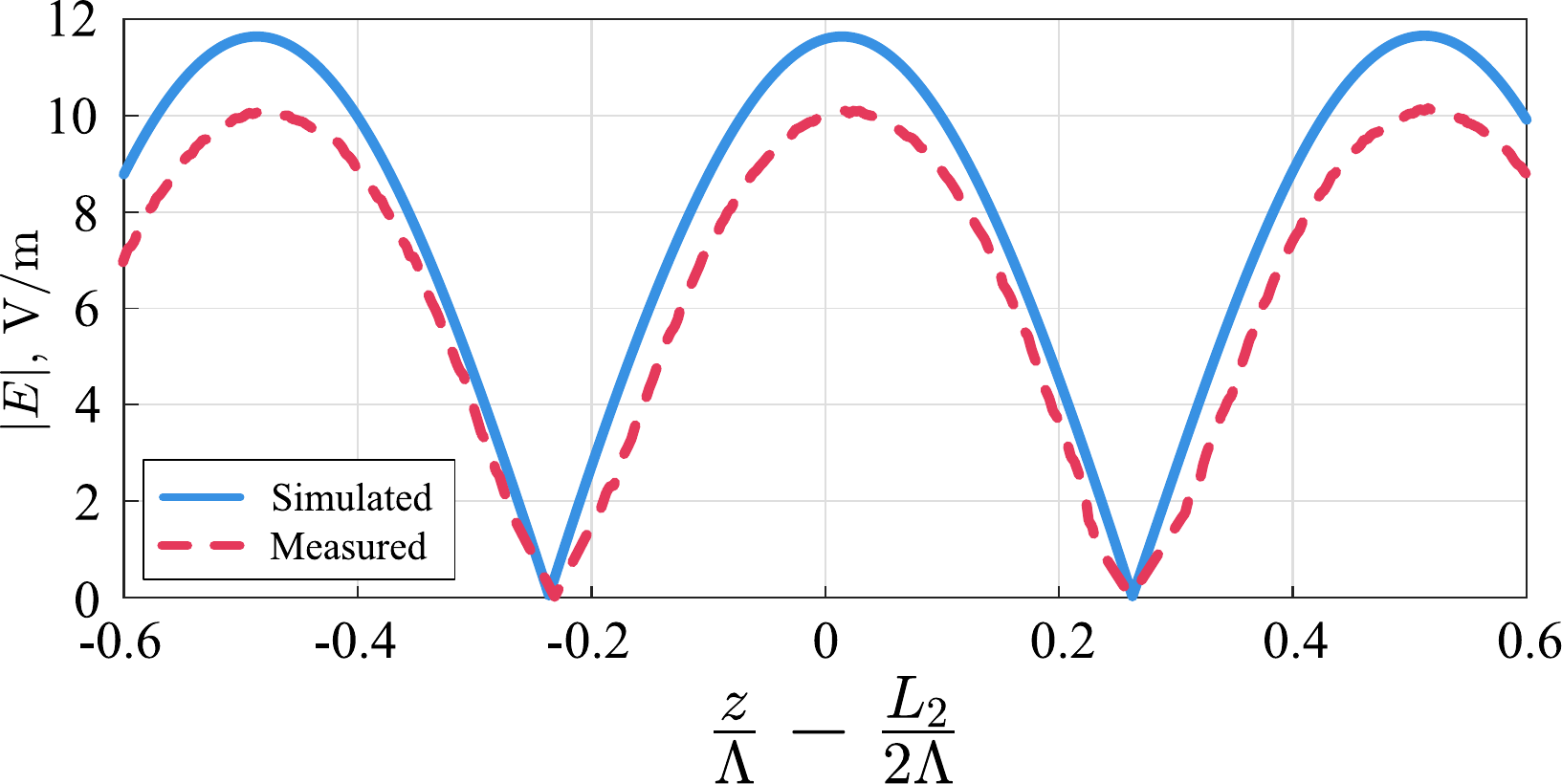} 
  \caption{Simulated and measured distributions of electric field magnitude across the central part of an invisible cavity composed of capacitive and inductive unit cells placed in a waveguide (the ninth order thickness resonance at $11.11$~GHz). Here $L_2=150$~mm is the distance between the metasurfaces in the second experiment.}
  \label{Field_waveguide}
\end{figure}
\begin{figure*}
 \center
  \includegraphics[width=0.95\textwidth]{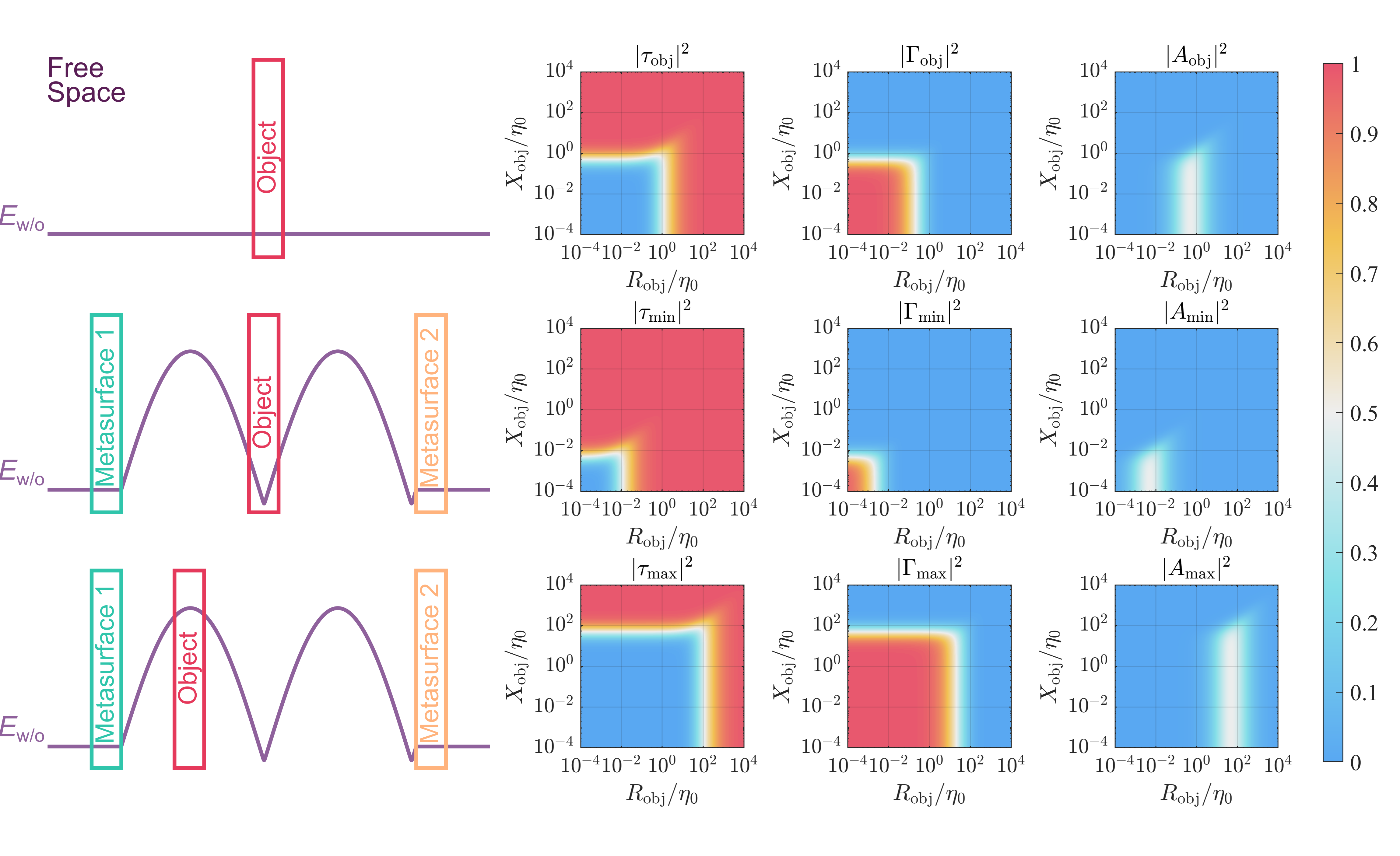}
  \caption{First, second, and third rows represent scenarios 
 of an object placed in free space, at the field minimum,  and at the field maximum of the cavity, respectively. The plots on the right depict transmittance $\vert \tau\vert^2$, reflectance $\vert \Gamma\vert^2$, and absorbance $\vert A\vert^2$ of incident waves for these scenarios. The cavity metasurfaces have reactances $X\@{e1}=-X\@{e2}=  38.08\,\Omega$. On the left side, $E_{\rm w/o}$ denotes the total electric field without the object.}
  \label{fig:sense_comp}
\end{figure*}

Next, to demonstrate the effect of resonant field enhancement in the invisible cavity, the same manufactured unit cells were employed, but the uniform section was replaced with another, three times longer calibrated measurement waveguide section P1-20 with the length  $L_2=150$~mm, as shown in Fig.~\ref{Setup}(b). The latter had a longitudinal thin slot in the middle of the broad side of the waveguide and was equipped with a movable thin-wire probe to measure the electric field distribution with minimum insertion and reflection losses. The probe was connected to the digital oscilloscope Rohde \& Schwarz HMO2022 through a diode detector. The waveguide was fed from one side by connecting the signal generator Rohde \& Schwarz SMB100A, while loaded with a matched load from the other side. The generator produced a wave at the carrier frequency of $11.11$~GHz modulated with a harmonic signal with the frequency of $100$~kHz. The probe was mechanically moved to plot the electric field magnitude versus the longitudinal coordinate by checking the amplitude of the detected signal at the modulation frequency and applying the calibration characteristic of the detector. 
Normalized to the field magnitude in the waveguide section without the cavity, the field can be determined quantitatively. 

For comparison, the realized metasurface unit cells were numerically simulated inside a waveguide section and the calculated field patterns were compared with the measured ones. The measured and simulated electric field distributions at the resonant frequency of $11.11$~GHz are compared in Fig.~\ref{Field_waveguide}, where the field in the empty waveguide was taken as $1$~V/m. 
Due to the physical constrains related with the slotted waveguide probe, only one third of the invisible cavity was characterized, as shown in Fig.~\ref{Field_waveguide} for the coordinate range from $-0.6 \Lambda$ to $0.6\Lambda$~mm.

The measured field profile is in good agreement with the simulated one. The difference of the measured resonant field enhancement   from the numerically predicted value (10 versus  $11.6$) can be explained by radiation leakage losses due to the measurement slot and the probe effect. Therefore, it was observed that the designed non-scattering cavity at $11.11$~GHz causes an order-of-magnitude field enhancement and suppression. Likewise, these results are in good agreement with theoretical predictions for lossy metasurfaces  (see Appendix~C under the assumption of $R_{\rm loss}/|X_{\rm e}| \approx 5\cdot 10^{-3}$).

\section{Applications of invisible cavities}
\subsection*{Manipulation of scattering from an object placed inside the cavity}
As was discussed above, the proposed cavities provide  enhancement and suppression of fields inside, keeping the fields outside undisturbed. Such functionality provides unique opportunities for manipulating scattered fields from objects placed inside the cavity. Indeed, placing an object in the field minima (maxima), one can suppress (enhance) wave scattering from it since the local field at the object position is decreased (increased) compared to the incident field. The best effect is achieved when the object is planar (since we consider 1D invisible cavities). 

%
%
%
%
%

%
We model  the object whose scattering should be enhanced or suppressed  as a thin sheet   with the  grid impedance $Z\@{obj}=R\@{obj}+j X\@{obj}$ (only electric polarization response is assumed). Positioning the object at a distance $z=\delta_{d}$, we can determine reflected and transmitted fields from it in the presence and absence of the cavity  (see Appendix~D). Figure~\ref{fig:sense_comp} shows the reflectance, transmittance, and absorbance of incident waves in three scenarios:  When the object is in free space (without the cavity), and when  it is placed in the field maximum and minimum inside the cavity (here, we chose $X\@{e1}=-X\@{e2}=  38.08\,\Omega$). 

Comparison of the plots in the first and second rows (free-space and field-minimum scenarios) reveals that by positioning an arbitrary object at the  field minimum of the cavity, one can drastically decrease reflection from the object. Absorption can be decreased or increased depending on the grid impedance of the object. The transmittance in this case is more  than $0.9$ when $|R\@{obj}+j X\@{obj}|>0.1\eta_0$. The phase of transmission is close to zero, which implies strongly suppressed   scattering from the object (note that the object alone    transmits less than 3\% when $|R\@{obj}+j X\@{obj}|=0.1\eta_0$). Thus, the invisible cavity can be exploited for scattering suppression from planar layers. 
The smaller grid reactances of the  metasurfaces, the greater suppression effect can be achieved (higher Q-factor). On the contrary, the smaller grid impedance  of the object to be hidden, the lower the effect. In the limit of the object made from perfect electric conductor, full reflection occurs even in the presence of the cavity.  
By comparing the first and third rows in Fig.~\ref{fig:sense_comp} (the free-space and field-maximum scenarios), one can  see that depending on the grid impedance of the object, the cavity can increase (up to 50\%) or decrease absorption in the object. At the same time, reflection is always increased and transmission decreased. Such functionality can be used for enhancement of the energy captured   by sensors. 

\begin{figure*}[t]
  \centering
  \includegraphics[width=1\linewidth]{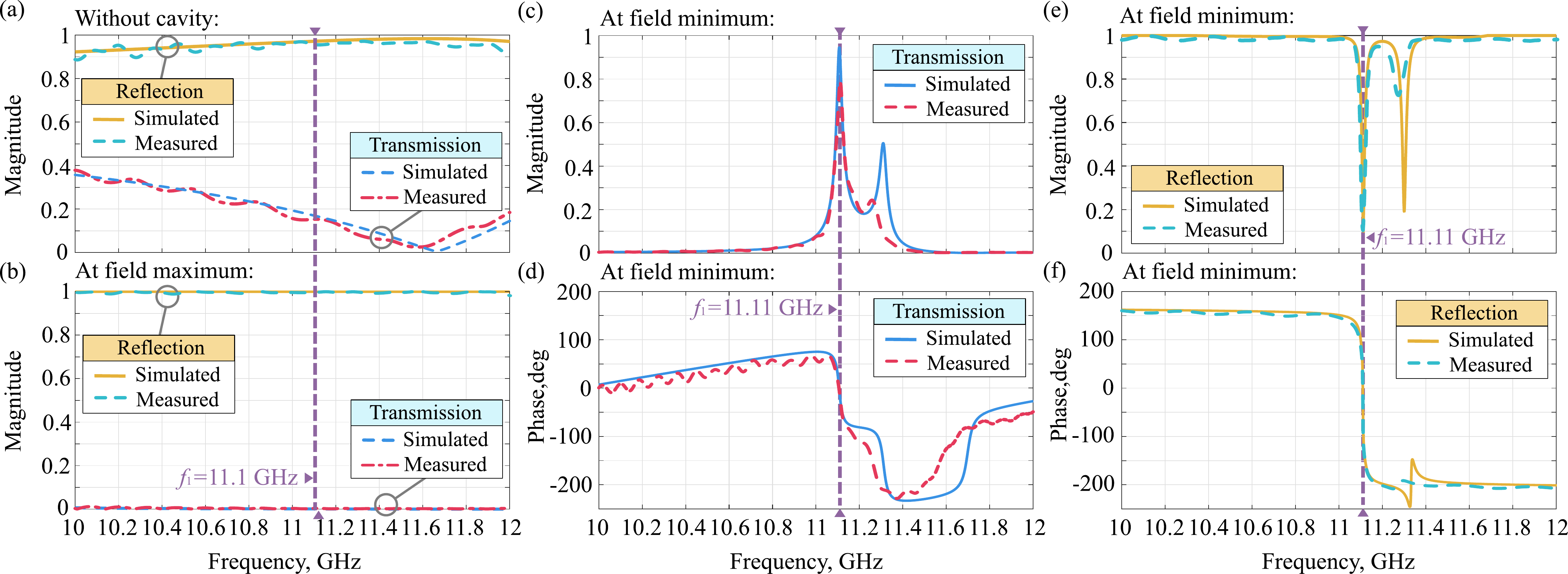} 
  \caption{Simulated and measured magnitudes of transmission and reflection coefficients: (a) Unit cell of the additional capacitive metasurface with the grid impedance of around $-j31~\Omega$ in the absence of the invisible cavity, (b) the same unit cell placed in the electric field maximum inside the designed  cavity, magnitude and phase of transmission (c,d) and reflection (e,f) coefficients for the additional capacitive metasurface placed in the minimum of the electric field of the designed cavity.}
  \label{Third_plate_meas}
\end{figure*}

To verify these results, we carried out the third experiment, where we measured  scattering from a planar object (with a  given grid impedance)  placed either at the minimum [Position~A indicated in Fig.~\ref{Setup}(c)] or at the maximum [Position~B] of the electric field distribution inside the cavity. 
To test scattering enhancement/suppresion, we used the same capacitive and inductive grids as shown in Fig.~\ref{Setup}(a) to form  the cavity. The planar object was also made as a printed circuit board with a carefully designed copper shape printed on one side of Arlon AD250 0.5-mm-thick substrate. However, to place the object into an arbitrary position in the previously used uniform section, the dimensions of the board were chosen almost equal to those of the waveguide cross section. At the same time, the metal pattern was designed to provide the grid impedance of $(1-j31)~\Omega$ at $f_1=11.11$~GHz with no electric contact to the waveguide walls required. 
The photograph of the object and the scheme of its insertion into the waveguide are illustrated on the insets in Fig.~\ref{Setup}(c). 
The shape was a rectangular patch of the dimensions  $L_{\text{i}x} \times L_{\text{i}y} = 14.2 \times 7.6$~mm$^2$, in which two symmetrical notches of the width of $0.2$~mm and the length of $4.1$~mm  were made to reach the desired grid capacitance. To fix the object  inside the waveguide, thin foam holders were employed.

As can be seen in Fig.~\ref{Third_plate_meas}(a), the  low grid capacitance of the object leads to high  reflection ($0.97$ at $11.11$~GHz) in the absence of the cavity. However, when the same object is placed in the electric field maximum inside the invisible cavity, the reflection coefficient becomes very close to unity in the whole frequency range from $10$ to $12$ GHz [as shown in Fig.~\ref{Third_plate_meas}(b)]. Therefore, in this case one observes broadband enhancement of reflectivity or scattering from an object placed inside the cavity. 

On the contrary, when the object is placed in the minimum of the electric field distribution inside the cavity, the simulated and measured transmission and reflection coefficients shown in Fig.~\ref{Third_plate_meas}(c-f) look drastically different. Indeed, as is seen from Fig.~\ref{Third_plate_meas}(c), the measured transmission coefficient magnitude  reaches $0.79$ at the resonance frequency of the  cavity ($11.11$~GHz) despite that the inserted metasurface itself was highly reflective. Interestingly, the transmission coefficient phase is still zero at this frequency. The experimentally achieved transmission coefficient magnitude of $0.79$ differs from the simulated value of $0.95$ due to small misalignments of the capacitive and inductive unit cells as well as the object with respect to the cross-section plane. 
Nevertheless, it is possible to conclude that when the reflective metasurface is properly inserted into the invisible cavity, scattering is nearly totally suppressed (some residue scattering occurs due to  absorption in metal parts). The other noticeable effect is the second resonance appeared at 11.32~GHz, which can be explained by the fact that the object paired with the capacitive metasurface of the cavity forms a new Fabry-Perot cavity. In this new cavity both walls have the same reactance of $-j31~\Omega$, so this new cavity is not transparent, as one can observe from the transmission and reflection magnitudes at $11.32$~GHz in Fig.~\ref{Third_plate_meas}(c,e). Moreover, because of its Fabry-Perot nature, this new cavity does not provide zero transmission phase at its resonance frequency as can be clearly seen from Fig.~\ref{Third_plate_meas}(d).

The above mentioned method to manipulate scattering is applicable for thin sheets with arbitrary grid impedances. However, if the  object impedance is known,   scattering can be suppressed  by cascading the object and one metasurface (with the opposite grid impedance). In this way, the object plays the role of the second metasurface in an invisible cavity and the total scattering from this system becomes zero.  

\subsection*{Transparent waveguides}
\begin{figure}
  \centering
  \includegraphics[width=1\linewidth]{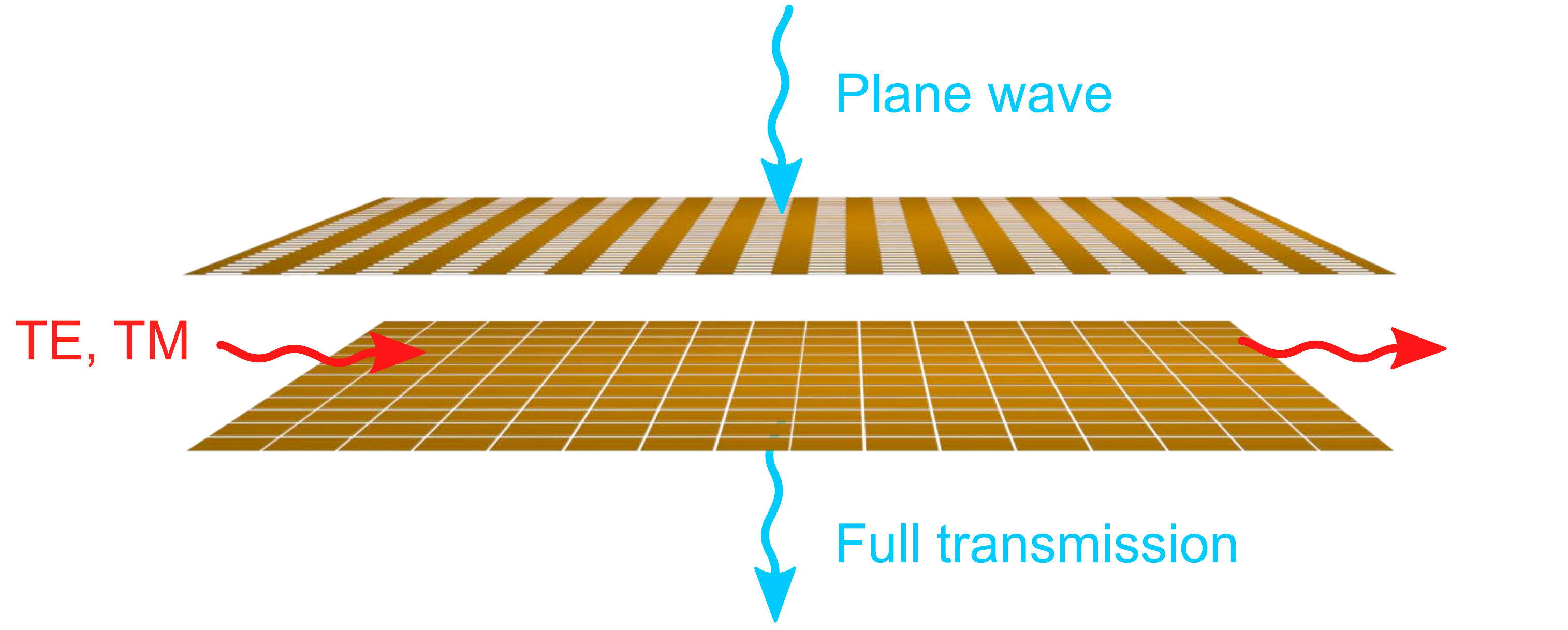} 
  \caption{An invisible cavity operates as a waveguide for transverse electric and magnetic waves which is completely transparent for normal-incidence illuminations at the operating frequency. }
  \label{transparent_waveguide}
\end{figure}

An invisible resonator made of high-reflective metasurfaces illuminated at normal incidence ensures field localization without scattering. On the other hand, a wave propagating parallel to the metasurfaces can be also localized between them, like in a parallel-plate waveguide \cite{Xin2019}. To consider this property, 
 we studied guided modes which can propagate between the metasurfaces for the same structure as studied above. At the frequency $f=11.2$~GHz, at which the transmission coefficient is unity for normal incidence (assuming $X_{\rm e1}=-X_{\rm e2}=38.08\,\Omega$, $d=\lambda_{\rm op}/2$), it is simultaneously possible to excite two different modes at the input port of the waveguide, as shown in Fig.~\ref{transparent_waveguide}: a transverse electric mode TE (whose longitudinal component of electric field is zero) and a transverse magnetic mode TM (whose longitudinal component of magnetic field is zero), respectively.
For both modes, the energy is confined inside the cavity and exponentially decays in free space outside the cavity. However, the attenuation constant is quite different for the two modes. The TE mode has a higher attenuation constant ($\alpha=1150$ m$^{-1}$) compared to the TM mode ($\alpha=472$ m$^{-1}$). Knowing the attenuation constants, we can find the distance $\delta$ away from the metasurface at which the field reaches $1 /\rm{e} $ of its magnitude at the metasurface boundary. For the TE mode  $\delta\approx0.87$~mm, while $\delta\approx2.12$~mm for the TM mode (less than 1/12 of the wavelength). 
It is worth noting that the phase and attenuation constants remain unchanged for different distances between metasurfaces.

\subsection*{Multi-layer resonators}

\begin{figure}
  \centering
  \includegraphics[width=1\linewidth]{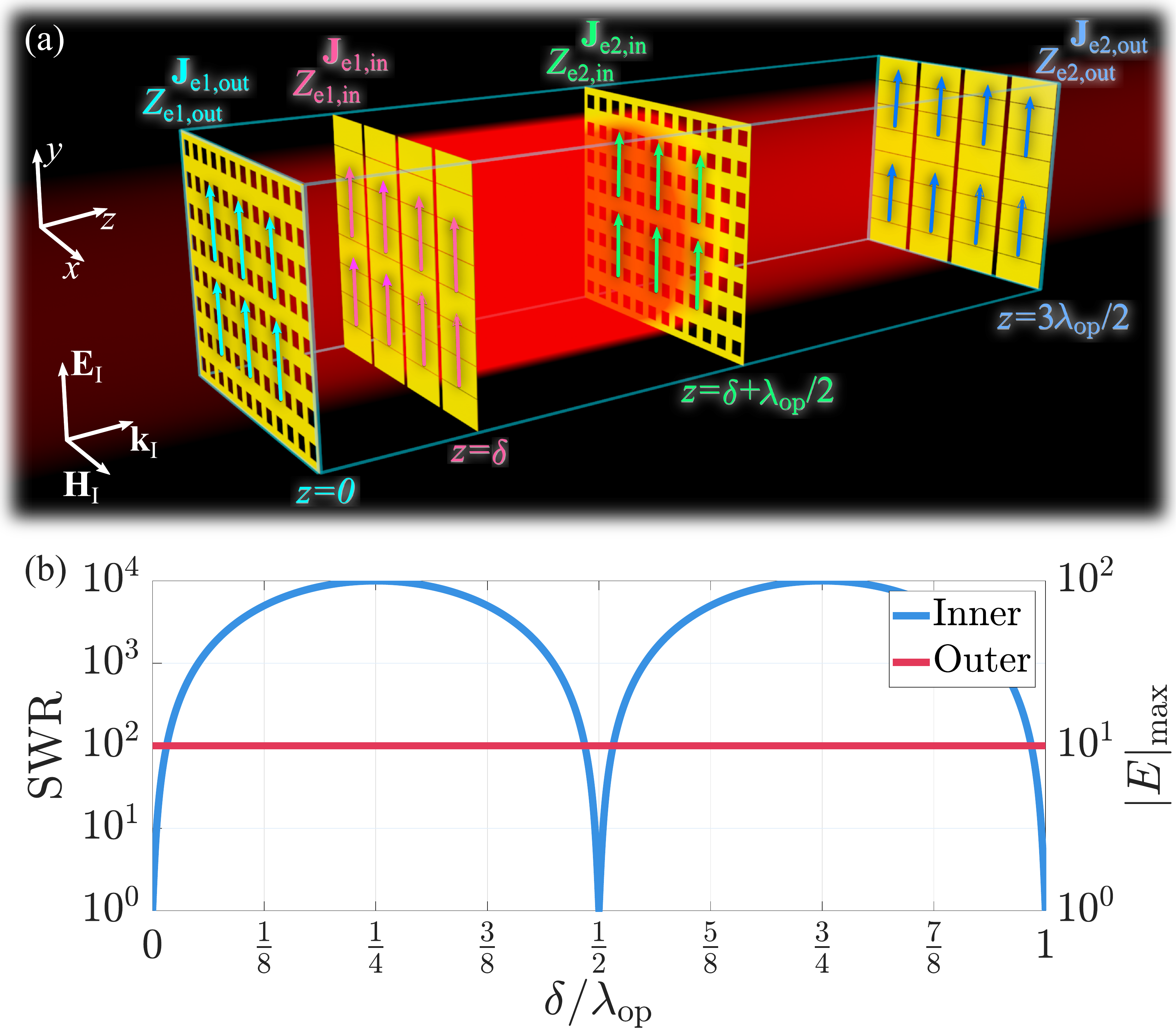} 
  \caption{(a) Geometry of the proposed multi-layer resonator obtained by inserting an invisible resonator of thickness $\lambda\@{op}/2$ at $z=\delta$ into a larger invisible resonator with thickness $3\lambda\@{op}/2$.  (b) SWR and the maximum electric field  of the standing waves produced inside the inner and outer cavities in a ``matryoshka''-like cavity configuration. The inner resonator has the distance between the metasurfaces of $\lambda_{\rm op}/2$, while  the outer resonator of $3\lambda_{\rm op}/2$. The parameter $\delta$ is the displacement of the inner resonator with respect to the origin (where the first metasurface of the outer resonator is located). The multi-layer resonator is composed by alternating layers of inductive and capacitive metasurfaces, such that extreme localization positions are located at multiples of $\lambda_{\rm op}/4$. The grid reactances of the metasurfaces were chosen as  $X_{\rm e1, out}=-X_{\rm e1, in}=X_{\rm e2, in}=-X_{\rm e2, out}=38\,\Omega$. }
  \label{matryoshka}
\end{figure}

An intuitive method to achieve even higher field localization is inserting an open cavity resonator inside another one [as shown in Fig.~\ref{matryoshka}(a)], such that the standing wave inside the inner cavity is fed by the outer resonator standing wave. Stacking open cavities which  produce non-zero   scattering, such as conventional Fabry-Perot resonators, is possible but not intuitive since it requires careful design of both cavities taking into account complex interference between them \cite{Stadt1985}. 
However, if one utilizes the invisible cavities proposed in this paper, the design becomes straightforward. Indeed, since the inner cavity generates no external scattering, the   standing wave in the outer cavity is not affected by it. Furthermore, the zero-scattering property of the combined  structure is granted regardless of the inner cavity position.
This peculiar combination of cavities, resembling a ``matryoshka'' doll, makes the displacement of the inner resonator $\delta$   an additional degree of freedom which can be exploited to achieve giant field enhancement inside the cavity and high SWR. The scattered fields across the structure can be found solving similar electromagnetic problems as done before for a single resonator [see Eqs.~\r{GSTC_gen} in Appendix~A], by just adding two more metasurfaces at $z=\delta$ and $z=\delta+n\lambda\@{op}/2$ and considering the inner and outer pairs as invisible resonators [using conditions of Eqs.~\r{eqfreespace}]. In the case that the multi-layer resonator is composed by alternating grid impedances $Z\@{e1,out}=-Z\@{e1,in}=Z\@{e2,in}=-Z\@{e2,out}=Z\@{e}$, the standing wave can be defined by its inner forward and backward components:
\begin{subequations}\label{eq:matryoshka_prop}
\begin{align}
	F\@{in}&=\dfrac{E\@{F,in}}{E\@{I}}=\dfrac{\left(e^{2 j k_0 \delta}-1\right){\eta_0}^2+4 Z\@{e}^2}{4 Z\@{e}^2}, \label{eq:mtk_fow}\\
	B\@{in}&=\dfrac{E\@{B,in}}{E\@{I}}=\frac{\eta_0(1-e^{-2 j k_0 \delta}) (2 Z\@{e}-\eta_0)}{4 Z\@{e}^2}, \label{eq:mtk_bak} \\
{\rm SWR}\@{in}&=\dfrac{\vert F\@{in}\vert+\vert B\@{in}\vert }{\vert F\@{in}\vert-\vert B\@{in}\vert }, \label{eq:mtk_swr}
\end{align}
\end{subequations}
These expression for SWR can be obtained using Eqs.~\r{s_swr} and \r{s_reflc} of Appendix~B. 

Figure~\ref{matryoshka}(b) demonstrates that by adjusting the distance $\delta$,  SWR and the maximal of electric field can be enhanced reaching the square values of those corresponding to a single cavity with given grid reactance (in this example we consider metasurfaces with grid impedance $Z_{\rm e}=j38\,\Omega$). On the other hand, the inner resonator can be positioned in such a way that there is no field enhancement at all. This invisible multi-layer configuration can be easily extended to any multiple of stacked cavities, which makes it very appealing for  applications where  high localization and field enhancement are required.  

\section{Discussion}
In this work, we have proposed and designed   one-dimensional cavities which are imperceivable (non-scattering) for an external observer  while still  provide strong field localization inside their volumes. The non-scattering regime occurs in a cavity formed by two metasurfaces due to destructive interference of waves scattered by each metasurface. 
In the former case, the distance between metasurfaces must be equal to an integer of half-wavelengths, while their grid impedances must have the opposite signs.   The non-scattering mode excited in such a cavity is driven by the incident wave and resembles an ideal bound state in the continuum of electromagnetic frequency spectrum. In contrast to known bound states in the continuum, the mode can stay localized in the cavity infinitely long without scattering, provided that the incident wave illuminates it. In the limit of zero reactances of the two sheets, the resonant mode converges to an ideal embedded eigenstate in the continuum. 

Although the described theory analyzes a one-dimensional scenario of invisible cavities, it can be extended in a similar way to the two- and three-dimensional scenarios. In fact, with appropriate modifications, the fabricated cavity was proved to be functional inside a rectangular waveguide, preserving all its unique properties. 
With the proper positioning of a thin planar object inside the cavity, one can increase or reduce electromagnetic scattering from it, with the potential applications in cloaking and sensing.
Moreover,  the proposed structures can be used to guide transverse electromagnetic waves along them, remaining  invisible for normally incident radiation (invisible parallel-plate waveguides).
Due to the reciprocity of the structure, the cavity offers new possibilities for emission control of active sources located inside it. 

\section*{Appendix A:\\ Solution of the boundary problem and conditions for the required grid impedances}

Consider the scenario illustrated in Fig.~\ref{fig:two_EP_fields}, where two metasurfaces are separated by a distance $d$ . Under normal plane-wave illumination by an incident wave with electric field $\_E\@{I}$,  the metasurfaces create secondary electromagnetic plane waves. The reflected (``R") and transmitted (``T") waves propagate in the $z<0$ and $z>d$ regions, respectively. 
Meanwhile, a standing wave inside the cavity can be described as a  superposition of  forward (``F")   and backward (``B") waves. The electric fields of these plane waves read 
\begin{subequations}\label{eq:teo_fields}
\begin{align}
	\_E\@{I}(z) = E\@{I} e^{-j k_0 z}\_a_{x},&\,\,\,\,\,\,\,\,\,\,\_H\@{I}(z) =\dfrac{E\@{I}}{\eta_0}  e^{-j k_0 z}\_a_{y}, \label{eq:e_inc}\\
	\_E\@{T}(z) = E\@{T} e^{-j k_0 z}\_a_{x},&\,\,\,\,\,\,\,\,\,\,\_H\@{T}(z) =\dfrac{E\@{T}}{\eta_0}  e^{-j k_0 z}\_a_{y}, \label{eq:e_trans}\\
	\_E\@{R}(z) = E\@{R} e^{j k_0 z}\_a_{x},&\,\,\,\,\,\,\,\,\,\,\_H\@{R}(z) =-\dfrac{E\@{R}}{\eta_0}  e^{j k_0 z}\_a_{y}. \label{eq:e_reflc}\\
	\_E\@{F}(z) = E\@{F} e^{-j k z}\_a_{x},&\,\,\,\,\,\,\,\,\,\,\_H\@{F}(z) =\dfrac{E\@{F}}{\eta_0}  e^{-j k z}\_a_{y}, \label{eq:e_fw}\\
	\_E\@{B}(z) = E\@{B} e^{j k z}\_a_{x},&\,\,\,\,\,\,\,\,\,\,\_H\@{B}(z) =-\dfrac{E\@{B}}{\eta_0}  e^{j k z}\_a_{y}, \label{eq:e_bw}
\end{align}
\end{subequations}
where $k_0$ and $\eta_0$ are the wavenumber and the characteristic impedance of the medium (assumed to be vacuum for simplicity).


Let us write the generalized sheet transition conditions for each metasurface:
\begin{subequations}\label{eq:GSTC_gen}
\begin{align}
	\_E\@{I}(0)+\_E\@{R}(0)&=\_E\@{F}(0)+\_E\@{B}(0), \label{eq:GSTC_e1}\\
	\_a\@{z}\times\left(\_H\@{F}(0)+\_H\@{B}(0)\right.&\left.-\_H\@{I}(0)-\_H\@{R}(0)\right)=\_J\@{e1}, \label{eq:GSTC_m1}\\
	\_E\@{F}(d)+\_E\@{B}(&d)=\_E\@{T}(d), \label{eq:GSTC_e2}\\
	\_a\@{z}\times\left(\_H\@{T}(d)-\_H\@{F}(\right.&\left.d)-\_H\@{B}(d)\right)=\_J\@{e2}, \label{eq:GSTC_m2}
\end{align}
\end{subequations}
where $\_J\@{e1}$ and $\_J\@{e2}$ are the electric surface current densities at each metasurface, respectively. The current density satisfies impedance boundary conditions, where the grid impedance $Z\@{e}$ is defined by the geometry and properties of the unit cells and the array period. In our scenario, substituting the expressions for the electric fields, the impedance relations take the form
\begin{subequations}\label{eq:MS_curr}
\begin{align}
	\_J\@{e1}&=\dfrac{\_E\@{I}(0)+\_E\@{R}(0)+\_E\@{F}(0)+\_E\@{B}(0)}{2Z\@{e1}}, \label{eq:curr_e1}\\
	\_J\@{e2}&=\dfrac{\_E\@{F}(d)+\_E\@{B}(d)+\_E\@{T}(d)}{2Z\@{e2}}. \label{eq:curr_e2}
\end{align}
\end{subequations}
The magnitude of each plane wave can be determined as a function of the grid impedances, the electric properties of the propagating medium, and the distance between the metasurfaces:
\begin{subequations}\label{eq:e_transfunc_fs}
\begin{align}
\Delta =& e^{2 j k_0d}(\eta_0+2Z\@{e1})(\eta_0+2Z\@{e2})-{\eta_0}^2,\label{eq:e_den_fs}\\
		\Gamma = \dfrac{E\@{R}}{E\@{I}}=&  {\Delta}^{-1} \left[{\eta_0}^2-2Z\@{e1}\eta_0-e^{2 j k_0d}(\eta_0+2Z\@{e2})\eta_0\right], \label{eq:e_reflc_fs}\\
	\tau = \dfrac{E\@{T}}{E\@{I}}=& {\Delta}^{-1} \left[ 4 e^{2 j k_0d} Z\@{e1} Z\@{e2} \right], \label{eq:e_trans_fs}\\
	F = \dfrac{E\@{F}}{E\@{I}}=& {\Delta}^{-1} \left[ 2 e^{2 j k_0d} Z\@{e1} (\eta_0+2 Z\@{e2}) \right], \label{eq:e_forw_fs}\\
	B = \dfrac{E\@{B}}{E\@{I}}=& {\Delta}^{-1} \left[-2 Z\@{e1} \eta_0\right]. \label{eq:e_back_fs}
\end{align}
\end{subequations}
Requiring the absence of scattering in Eqs.~\r{e_transfunc_fs} ($\Gamma=0$ and $\tau=1$), the conditions of Eqs.~\r{eqfreespace} are met.

\section*{Appendix B:\\ Properties of Invisible Cavities}
\subsection*{Fields inside the cavity and the standing-wave ratio}

\begin{figure}[tb]
    \centerline{\includegraphics[width=80mm]{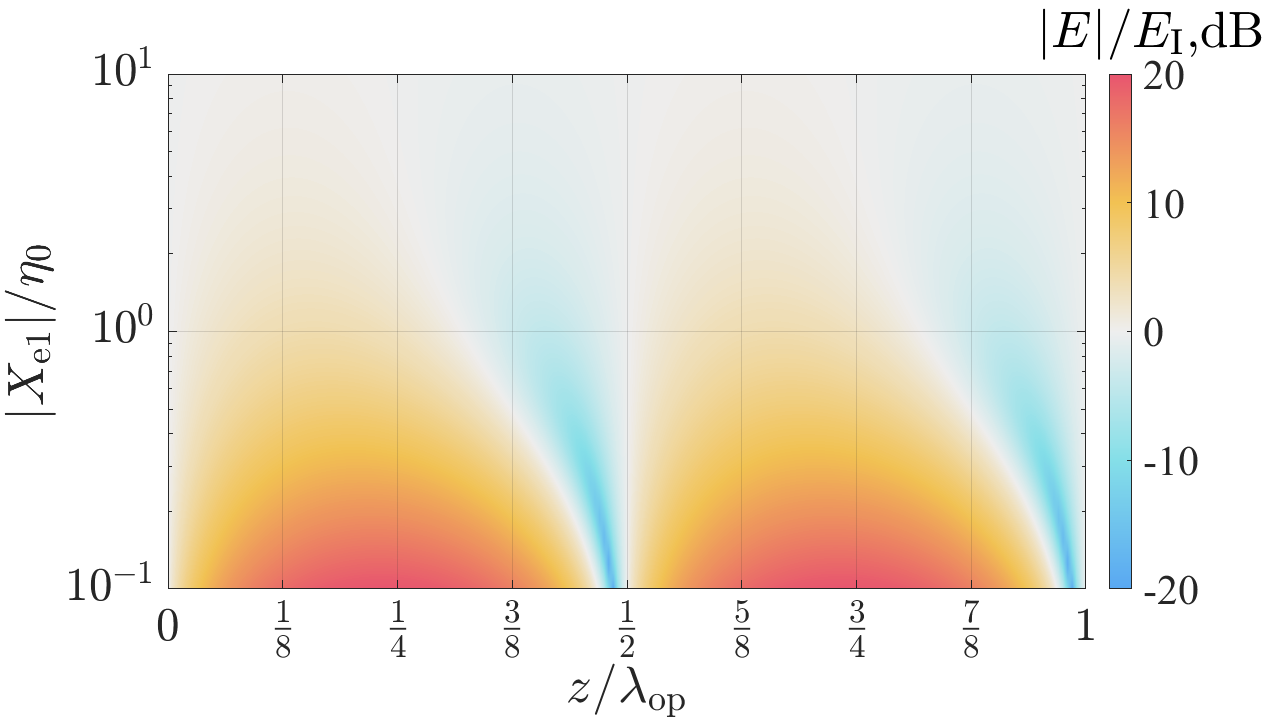}}
    \caption{Magnitude of the total electric field  across a lossless invisible cavity for different reactance  values. The cavity  thickness is $d=\lambda_{\rm op}$. Notice that the color bar is in dB scale.}
    \label{fig:teo_performance2}
\end{figure}

Plane waves created by metasurface currents can form a standing wave between them. The total internal electric field $\_E$ is the sum of the forward wave $\_E\@{F}$ and the backward wave $\_E\@{B}$:
\e \begin{array}{c}
\_E(z)= \_E\@{F}(z)+\_E\@{B}(z)=\\\displaystyle
\dfrac{E\@{I}}{2Z\@{e1}}\left[\left(2Z\@{e1}-\eta_0\right)e^{-jk_0z}-\eta_0e^{jk_0z}\right]\_a\@{x}. 
\end{array}\l{s_e_field}\f
Figure~\ref{fig:teo_performance2} illustrates the electric field strength at different locations inside a lossless cavity ($Z_{\rm e 1,2}=jX_{\rm e 1,2}$, $d=\lambda_{\rm op}$) as a function of the reactance value $X_{\rm e1}$. As is typical for standing waves, it is seen that the field distribution has a repeating pattern along the $z$ direction with the periodicity of $\lambda_{\rm op}/2$. Each period includes points where the electric field is  minimum and maximum. As the impedance values $|X_{\rm e 1,2}|$ decrease (metasurfaces behave more similar to a perfect electric conductor sheet), the total field inside the cavity increases. 

The simplest method to characterize field localization is through the Standing Wave Ratio (SWR), which is the ratio between the maximum and the minimum values of the amplitude of the total field \cite{Pozar2011}. The SWR is defined as
\e \rm SWR =\dfrac{\vert E\vert\@{max}}{\vert E\vert\@{min}} = \dfrac{1+\vert\Gamma_{\rm int}\vert}{1-\vert\Gamma_{\rm int}\vert},\l{s_swr} \f
where $\Gamma\@{int}$ corresponds to the internal reflection coefficient:
\e\Gamma_{\rm int}=\dfrac{\_E\@{B}\cdot\_a_{x}}{\_E\@{F}\cdot\_a_{x}}=\dfrac{B}{F}.  \l{s_reflc}\f
Using the above result, the SWR can be written in an elegant form of Eqs.\r{SWR}.

The second characteristic of partially standing waves is the positions of their maxima and minima $z_{0}$, which can be found using the first and second derivatives of the magnitude of the total field $\_E$ given by Eqs.~\r{s_e_field}. Due to mathematical complexity of the solution, positions of the maxima and minima were obtained  only for the scenario of purely reactive metasurfaces  ($Z\@{e1}=-Z\@{e2}=jX\@{e1}$):
\e z_{0}=\dfrac{\lambda\@{op}}{4}\bigg(p-\dfrac{\rm{arctan}(2\it X\@{e1}/\eta_0)}{\pi}\bigg), \label{eq:s_e_maxmin}\f
where $p$ is an integer, which denotes the property that consecutive maximum and minimum values are separated by $\lambda\@{op}/4$. The grid impedance value affects the spatial positions of the maxima and minima, as near-PEC impedances localize maxima/minima points to multiples of quarter-wavelength. Likewise, the sign of the reference grid impedance $Z\@{e1}$ determines the nature of the closest extreme value (maximum/minimum) to the origin.

\subsection*{Quality factor of the cavity}
As for any resonator, the quality factor can be used to characterize the resonant  properties of the cavity in the frequency domain. The Q-factor is defined as the ratio of the resonance frequency and the half-power bandwidth with respect to the structure transmittance:
\e Q=\dfrac{\omega\@{c}}{\Delta \omega},\,\,\,\,\,\, \vert\tau (\omega\@{c} \pm \Delta \omega /2)\vert^{2}=\dfrac{1}{2} \vert\tau\vert_{\rm max}^{2}, \l{qfactor_qdef}\f
where the half-power bandwidth ($\Delta \omega$) is defined as the frequency range where the transmittance is at least half of its maximum value (measured at the center angular frequency $\omega\@{c}$), which in the case of ideal invisibility is equal to unity. Therefore, the Q-factor can be found by replacing the transmittance in Eqs.~(\ref{eq:e_trans_fs}) at the cut-off frequencies $\omega\@{c}\pm\Delta\omega/2$, assuming frequency independent grid impedances of the metasurfaces. This leads to the result
\e \vert e^{2 j k_{\Delta} d}\left({\eta_0}^2-4Z\@{e1}^{2}\right)-{\eta_0}^2 \vert ^{2} = 32 \vert Z\@{e1}^{2} \vert ^{2},  \l{qfactor_qexact1} \f
where the variable $k_{\Delta}$ represents the wavenumber at the cut-off frequencies: $k_{\Delta}= k_0 \pm (\Delta \omega \sqrt{\varepsilon_0 \mu_0})/2$. If the phase product $k_{\Delta} d$ is considered, the expression can be reduced to $k_{\Delta} d = n\pi \left(1 \pm {\Delta \omega}/ 2\omega_0 \right)$. The first component of the sum will vanish after substituting it into the exponent in Eqs.~\r{qfactor_qexact1}, while the second component will give us the quality factor. The solution of Eqs.~\r{qfactor_qexact1} requires an expansion of the grid impedance ($Z\@{e1}=R\@{e1}+jX\@{e1}$) and the complex exponential $e^{2j k_{\Delta} d}$. This result can be simplified considerably for purely reactive metasurfaces  ($R\@{e1}=R\@{e2}=0$), which yields
\e \begin{array} {c}
32 X\@{e1}^{4} = \left[{\rm cos}\left(\dfrac{n\pi}{Q}\right)\left({\eta_0}^2+4\tilde{X}\@{e1}^{2}\right)-{\eta_0}^2\right]^{2}\\\displaystyle
+\left[{\rm sin}\left(\dfrac{n\pi}{Q}\right)\left({\eta_0}^2+4X\@{e1}^{2}\right)\right]^{2}. 
\end{array}\l{qfactor_qexact3} \f
This equation can be easily solved, resulting in 
\e Q=\dfrac{n\pi}{{\rm arccos} \left( 1- \dfrac{8X\@{e1}^{4}}{\eta_{0}^{4}+4{X}\@{e1}^{2}\eta_{0}^{2}} \right)}. \label{eq:qfactor_qexact}\f
For small values of the grid reactance ($\vert X\@{e1}\vert \ll \eta_0$), some approximations related with the sine function can be exploited and the quality factor can be reduced into the expression shown in Eqs.~\r{qfactor_redux}.

\section*{Appendix C:\\ Resistance evaluation for a lossy metal cladding}

\begin{figure*}[bt]
 \center
  \includegraphics[width=0.9\textwidth]{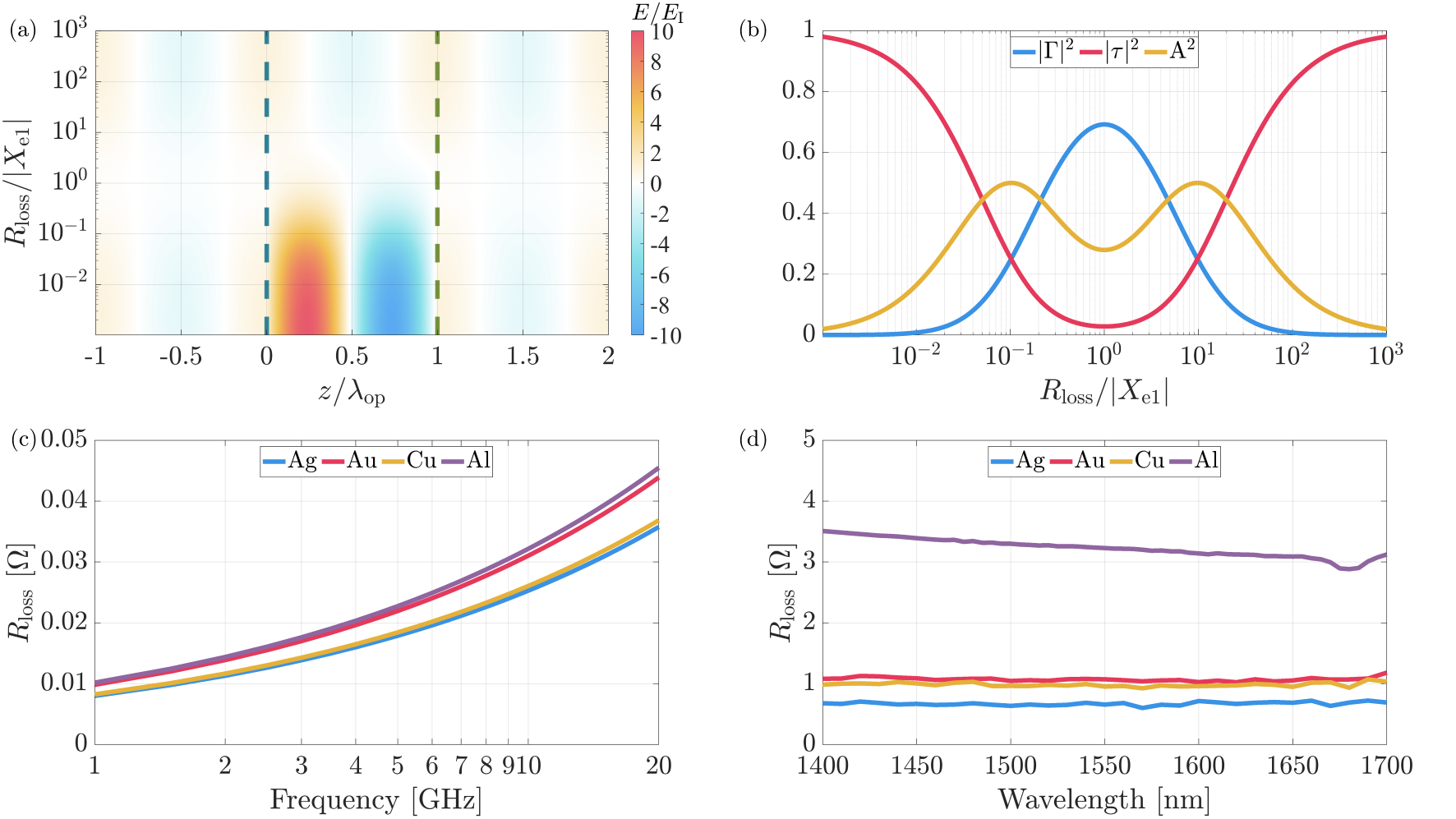}
  \caption{Analysis of the cavity performance in the presence of dissipation loss. The imaginary parts of the metasurface grid impedances was chosen $X\@{e1}=-X\@{e2}= 38.08\,   \Omega  $. (a) Effect of loss in metasurfaces   on the total electric field across the cavity. (b) Transmittance $\vert \tau\vert^2$, reflectance $\vert \Gamma\vert^2$, and absorbance $\vert A\vert^2$ through a cavity versus the normalized resistance of the metasurfaces. (c) Grid resistance evaluation of a continuous metal layer of~$ 17\,{\rm \mu m}$ thickness in the microwave frequency range. Data for different metals (silver, gold, copper, and aluminum) is shown. (d) Same as (c) but for the near infrared region. The metal layer thickness is   $ 30\,{\rm nm}$. \label{fig:losses}}
\end{figure*}

Previous analysis considered only ideal lossless metasurfaces, which is unrealistic in practice as losses affect performance of every device to some degree. Therefore, it is important to analyze how the presence of dissipation loss in metasurfaces will affect the performance of the invisible cavity. Naturally, loss can be compensated in the cavity via additional energy pumping which ensures that   condition~\r{eqfreespace} holds. However, in most applications, external energy pumping is not desired due to additional complexities. Therefore, we analyze the performance degradation  of cavities when the impedances of both metasurfaces have additional and equal positive real parts  $R\@{loss}$ corresponding to grid resistances:
\begin{equation} 
Z\@{e1}=Z\@{e2}^{*}=R\@{loss}+jX\@{e1}. \label{eq:z_loss}
\end{equation}
Taking into account the dissipation loss, the total electric field across the cavity is modified, as shown in  Fig.~\ref{fig:losses}(a).
If the grid losses are much greater than the reactances (in the absolute values), both metasurfaces are weakly excited, resulting in low scattering and low field amplitude inside the cavity.  When  the metasurface resistance is comparable to the reactance,
the cavity  becomes  reflective and also absorption   increases,  
as is seen   in Fig.~\ref{fig:losses}(b) which depicts the reflectance $|\Gamma|^2$, transmittance $|\tau|^2$, and absorbance $|A|^2$ through the cavity. The structure  remains nearly non-scattering only  when the ratio of the grid reactance and  resistance is sufficiently  small.  
Therefore, it is important to determine how small ratios $R\@{loss}/|X\@{e1}|$ can be achieved with realistic materials in different frequency ranges. Here, we analyze realizations in  the microwave  (wavelength from 1 to 20~GHz) and infrared (wavelength from 1.4 to 1.7~$\mu$m) regions. We evaluate grid resistance  $R\@{loss}$ under assumption that  each  metasurface represents a continuous layer of metal (gold, silver, copper, or aluminum) with a specific thickness $\Delta_{d}$. 
Naturally, metasurface patterning will result in  higher (for inductive metasurfaces) or lower (for capacitive metasurfaces) values of grid resistance compared to the  estimated ones. Nevertheless, such estimation gives a straightforward way to predict the influence of dissipation loss in  invisible cavities. 

The metallic layer can be modelled as a dielectric slab (with complex permittivity), which transmission and reflection coefficients are obtained by solving the boundary conditions between the slab and the surrounding medium [similarly as done before for the invisible resonator in Eqs.~\r{GSTC_gen}]:
\begin{subequations}\label{eq:loss_e_slabts}
\begin{align}
	\tau\@{loss} &= \dfrac{4 e^{j (k_{0}-\gamma) \Delta_{d} } \eta \eta_{0}}{\left(\eta+ \eta_{0} \right)^2-e^{-2\gamma \Delta_{d}}\left(\eta - \eta_{0} \right)^2},\label{eq:loss_e_slabtr}\\
	 	\Gamma\@{loss} &= \dfrac{\left(1-e^{-2 \gamma \Delta_{d} }\right) \left({\eta}^2 -{\eta_{0}}^2\right)}{\left(\eta+ \eta_{0} \right)^2-e^{-2\gamma \Delta_{d}}\left(\eta - \eta_{0} \right)^2}, \label{loss_e_slabrlc}
\end{align}
\end{subequations}
where $\gamma$ and $\eta$ correspond to the complex wavenumber and characteristic impedance of the metallic layer. In the case that the metallic layer does not exhibit  magnetic polarization (under the condition $\tau\@{loss} \approx 1+\Gamma\@{loss}$), the metallic layer can be replaced by a metasurface with similar transmission and reflection coefficients. The equivalent grid impedance (whose resistance $R\@{loss}$ models dissipation effects) is obtained from the analysis of a single metasurface:
\e Z\@{loss}=\dfrac{\tau\@{loss} \eta_0}{2\left(1-\tau\@{loss}\right)}=-\dfrac{\eta_0}{2}\left(\dfrac{1}{\Gamma\@{loss}}+1\right).\l{loss_e_slab_imp}\f

In the microwave range, metals can be considered as good conductors with
\begin{equation}
	\gamma=(1+j)\sqrt{\dfrac{2 \pi f\@{op} \mu\sigma_e}{2}},\,\,\,\,\,\eta=(1+j)\sqrt{\dfrac{2 \pi f \mu}{2\sigma_e}},\label{eq:slab_microwave}
\end{equation} 
where $\sigma_e$ is the metal conductance, $f\@{op}$ is the operational frequency and $\mu$ is the metal permeability (being equal to $\mu_0$, vacuum's permeability, for the selected metals) \cite{Pozar2011}. In this analysis, the assumed slab thickness  is    $17\,[{\rm \mu m}]$ (around $0.05\% \lambda\@{op}$ at 9~GHz), using metal conductivities  from Ref.~\cite{Pozar2011}.
The analytical results in Fig.~\ref{fig:losses}(c) demonstrate that in the microwave region the dissipation loss in metasurfaces is very small   and can be neglected  unless the    ratio $R\@{loss}/|X\@{e1}|$ is not small. Using Fig.~\ref{fig:losses}(b), one can estimate the minumum reactance $|X\@{e1}|$ which one can achieve without compromising the cavity operation. For example, at a frequency of 11~GHz, one can design copper invisible cavity  with reactance $|X\@{e1}|\approx 14\Omega$, providing that the transmittance will be not less than 90\%. Such small reactance ensures cavity Q-factor of the order of $10^3$, as is seen from Fig.~\ref{fig:teo_performance}(b).

Metals cannot be considered as good conductors at near-infrared frequencies, therefore, propagation parameters are written in a more way as
\begin{equation}
	\gamma=j 2 \pi f \sqrt{\mu \varepsilon },\,\,\,\,\,\eta=\sqrt{\dfrac{\mu} {\varepsilon} }.\label{eq:slab_optic}
\end{equation} 
For this analysis, we choose the slab thickness $\Delta_{d}=30\,{\rm nm}$ (approximately $2\%$ of the wavelength $\lambda=1550\,{\rm nm}$). Using metal parameters from Ref.~\cite{McPeak2015}, it was found that dissipation in metals  is high and cannot be neglected  [see Fig.~\ref{fig:losses}(d)]. In order to achieve high-Q invisible cavities in this range, one should use plasmonic metasurfaces with external energy pumping~\cite{Krasnok2018} or  design   invisible cavities based on high-permittivity dielectric metasurfaces~(e.g.,~\cite{Genevet2017}). In the latter case, one can envisage realization of metasurfaces using arrays of subwavelength dielectric spheres or cylinders operating near the electric  resonance. One metasurface operates at the frequency slightly below the resonance (negative $X_{\rm e1}$) and the other one slightly above the resonance (positive $X_{\rm e2}$). 

\section*{Appendix D: \\ Reflection and transmission through the cavity with an arbitrary object placed inside}
\begin{figure}[!t]
    \centerline{\includegraphics[width=0.45\textwidth]{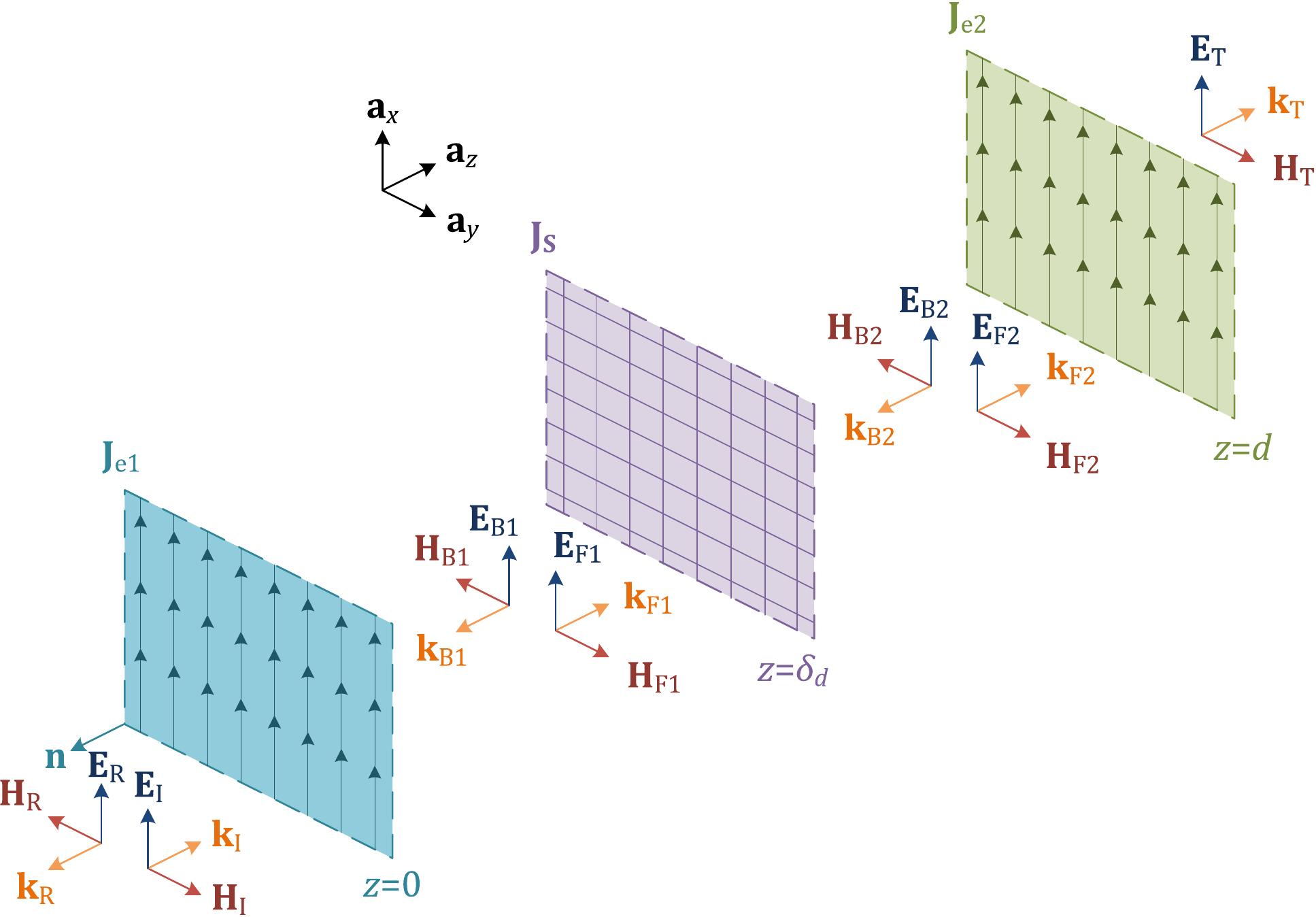}}
    \caption{Electromagnetic waves produced by normal illumination onto an invisible cavity with a planar object inside. \label{fig:sensor_e_meta_fields} }
\end{figure}

Consider the scenario of a scattering object positioned inside an invisible resonator at  $z=\delta_d$, as shown in Fig.~\ref{fig:sensor_e_meta_fields}. Without loss of generality, the object is modelled as a metasurface whose electric current density is given by the grid impedance $Z\@{obj}=R\@{obj}+jX\@{obj}$ and the average electric field at its plane.
Under these circumstances, the generalized sheet transition conditions are similar to those found in Eqs.~(\ref{eq:GSTC_gen}), with the addition of the discontinuities produced by the object:
\begin{subequations}\label{eq:GSTC_sense}
\begin{align}
	\_E\@{F1}(\delta_d)+\_E\@{B1}(\delta_d)&=\_E\@{F2}(\delta_d)+\_E\@{B2}(\delta_d), \label{eq:GSTCsense_eobj}\\
	\_a\@{z}\times\left(\_H\@{F2}(\delta_d)+\_H\@{B2}(\delta_d)\right.&\left.-\_H\@{F1}(\delta_d)+\_H\@{B1}(\delta_d)\right)=\_J\@{obj}, \label{eq:GSTCsense_mobj}
\end{align}
\end{subequations}
where the forward and backward propagating waves in front of and behind the object are denoted with subindices ``1'' and   ``2'', respectively; and $\_J\@{obj}$ is the electric current density induced in the object, defined as
\e \_J\@{obj}=\dfrac{\_E\@{F1}(\delta_d)+\_E\@{B1}(\delta_d)+\_E\@{F2}(\delta_d)+\_E\@{B2}(\delta_d)}{2Z\@{obj}}.\l{obj_curr}\f

The  boundary conditions, after applying invisibility conditions compatible with Eqs.~(\ref{eq:eqfreespace}), lead to the following transfer functions:
\begin{subequations}\label{eq:sensor_e_transfunc}
\begin{align}
	\tau=\dfrac{E\@{T}}{E\@{I}}=& -\Delta^{-1}8Z\@{e1}^2 e^{2 j k_0 \delta_d} Z\@{obj}, \label{eq:sensor_e_trans}\\
	\Gamma=\dfrac{E\@{R}}{E\@{I}}=&\Delta^{-1}\left[2 e^{2 j k_0 \delta_d} {\eta_0}^2(2 Z\@{e1}-{\eta_0})\right.\nonumber\\
	&\left.+e^{4 j k_0 \delta_d}{\eta_0}^3 +{\eta_0}({\eta_0}-2 Z\@{e1})^2\right],\label{eq:sensor_e_reflc}\\
    \Delta=&	2 e^{2 j k_0 \delta_d} \left[{\eta_0}^3 -2Z\@{e1}^2 (2 Z\@{obj}+{\eta_0}) \right]\nonumber\\
		&+{\eta_0}^2\left[(2 Z\@{e1}-{\eta_0})-e^{4 j k_0 \delta_d}(2 Z\@{e1}+{\eta_0})\right]. \label{eq:sensor_e_delta}
\end{align}
\end{subequations}
The results of Fig.~\ref{fig:sense_comp} were obtained by calculating transmission and reflection coefficients of Eqs.~(\ref{eq:sensor_e_transfunc}) for an object with variable grid impedance $Z\@{obj}$ located at the field maximum/minimum given by Eqs.~(\ref{eq:s_e_maxmin}). As the two metasurfaces forming the invisible resonator are lossless, the power absorbed in the object can be found as 
\e \vert A\vert^2=1-\vert \tau\vert^2-\vert \Gamma\vert^2.\l{abs_est}\f
\bibliographystyle{IEEEtran}
\bibliography{paper_ref}

\end{document}